%% file: main.tex
\documentclass[journal]{IEEEtran}
\usepackage{graphicx}
\usepackage{subfig}
\usepackage{xcolor}
\usepackage{xspace}
\usepackage[export]{adjustbox}
\usepackage{url}
\usepackage[acronym,nogroupskip]{glossaries}
\usepackage{multirow}
\usepackage{comment}
\usepackage{amsmath,amssymb,amstext}
\usepackage{soul}
\usepackage{cite}
\usepackage{enumitem}

\newcommand{\joao}[1]{\textcolor{blue}{#1}}

\title{Performance Analysis of Zero-Forcing Beamforming Strategies for the Uplink of an MU-MIMO System with Multi-Antenna Users}

\author{João Paulo P. G. Marques, Catherine Rosenberg \IEEEmembership{Fellow, IEEE}
\thanks{This work was conducted when all authors were at the Department of Electrical and Computer Engineering at the University of Waterloo, Canada.}}

\input{glossary.tex} 
\begin{document}

\maketitle
\begin{abstract}
We conduct a comprehensive evaluation of the performance of the uplink of  OFDMA-based MU-MIMO  systems with multi-antenna users, for three \gls{zf} \gls{bf} strategies: \gls{ctrone}, where only the strongest data stream is enabled per scheduled user; \gls{bd}, where all possible streams are enabled per scheduled user; \gls{ctrf}, which allows a flexible stream allocation per user.
The Radio Resource Management (RRM) of the uplink of all OFDMA-based systems must be done over an entire \gls{ts} due to power management, making it challenging. To enable this study, we propose an efficient heuristic based on  greedy-up searches for stream-sets that provides feasible solutions. It operates over the TS and considers fairness, practical Modulation and Coding Schemes and all RRM processes.  The results show that, for \glsentrylong{rma} scenarios,  \gls{bd} (resp. \gls{ctrone}) could replace the more complex \gls{ctrf} if the number of users is small (resp. large), while for \glsentrylong{uma} scenarios, \gls{ctrone} emerges as an alternative to \gls{ctrf} due to its similar performance. We also show that the system parameters can substantially impact the performance of the ZF strategies and that  \gls{bd} performance is more impaired  with a simpler power management scheme than \gls{ctrone} and \gls{ctrf}. 
\end{abstract}


\glsresetall
\section{Introduction} \label{sec:intro}

\IEEEPARstart{M}{ulti}-user Multiple-Input Multiple-Output (\glsdisp{mu-mimo}{MU-MIMO}) is a key technology to enhance the performance of wireless cellular systems by allowing transmissions of multiple data streams at the same time and frequency~\cite{Marzetta-2015}. This work considers the \gls{ul} of an \gls{mu-mimo} single-cell system with \gls{ma} users and \gls{ofdma}. Most of the studies on \gls{mu-mimo} have focused on \gls{sa} users and the \gls{dl} \cite{Castaneda-2017}. With \gls{ma} users, there is greater potential for performance improvement as the multiple antennas can be used to enable one strong data stream for many users or several data streams per user for fewer users~\cite{Bjornson-2013}.
Addressing the \gls{ul} has become increasingly critical given the growth in the \gls{ul} traffic and the \gls{ul} impact on the perceived quality of service \cite{Haseen-2024, Ozcan-2021}. 

\gls{rrm}  must be carefully addressed to achieve the \gls{mu-mimo} potential both on the DL and on the UL. \gls{rrm} with MA users is composed of the following inter-related processes:   \gls{ss} (as opposed to User Selection with \gls{sa} users), \gls{bf}, power management and \gls{mcs} selection. Power management with \gls{ma} users comprises \gls{pa} and \gls{pd} on both the \gls{ul} and the \gls{dl} as opposed to \gls{pd} (resp. \gls{pa}) alone for the \gls{dl} (resp. \gls{ul}) for \gls{sa} users.

\gls{ss} is the process that selects a subset of streams for transmission in each \gls{prb}\footnote{A pair made of a subchannel and a time-slot is the smallest scheduling unit and is called a \gls{prb}.}. On the \gls{dl}, \gls{ss}  can be done independently on each \gls{prb}, while on the \gls{ul}, it has to be done on all the \glspl{prb} within a \gls{ts} as will be explained shortly.

\gls{bf} is the process that enables the \gls{bs} to send to or receive from multiple users at once on the \gls{dl} or the \gls{ul}, respectively. \gls{zf} \gls{bf} comprises a family of techniques that eliminates inter-user interference \cite{Boccardi-2007, Spencer-2004}. They also enable the decoupling of \gls{bf} and power management, simplifying the RRM procedure. \gls{zf} \gls{bf} techniques have been studied extensively in the literature and several efficient computational methods have been developed~\cite{Wang-2008, Shi-2008, Tran-2012-1, Tran-2012-2}. They are the \gls{bf} techniques used in this paper. There are three primary \gls{zf} \gls{bf} strategies based on the number of data streams used per scheduled user: 1) \gls{ctrone} that uses exactly one stream per selected user (the strongest)~\cite{Shi-2008}; 2) \gls{bd} that uses all possible streams per selected user~\cite{Spencer-2004} (this number depends on the number of antennas at the user); 3) \gls{ctrf} that allows flexible stream allocation~\cite{Boccardi-2007}.
\gls{ctrf} can deliver better performance as any stream combination can be enabled per user at the price of higher \gls{bf} computational complexity compared to the other strategies~\cite{Boccardi-2007, Spencer-2004}.  Also, the \gls{ss} process under \gls{ctrf} is more complex due to the flexibility in the stream allocation, while for \gls{ctrone} or \gls{bd}, the stream allocation is fixed for each scheduled user~\cite{Chen-2008}. Investigating offline (i.e., without time constraints) when the simpler strategies \gls{ctrone} and \gls{bd} perform comparably to \gls{ctrf} is paramount to the design of real-time \gls{rrm} solutions for \gls{mu-mimo} systems with \gls{ma} users. While this was studied on the \gls{dl} in \cite{Bjornson-2013, Boccardi-2007, Shi-2008, Joao-2025, Shen-2006}, there is no equivalent study on the \gls{ul}. In fact, there are no studies on \gls{ul} \gls{rrm} for \gls{ctrf} to the best of our knowledge and only very limited studies for \gls{bd}.

The \gls{ul} \gls{rrm} problem is more challenging than its \gls{dl} counterpart because of power management. On the \gls{ul}, the power belongs to each user individually and the \gls{pa} process allocates the power budget in a given \gls{ts} across the \glspl{prb} for which the user is selected for transmission. Then, with \gls{ma} users,  \gls{pd}  is required for distributing the power allocated to each \gls{prb} by \gls{pa} among the user’s data streams enabled in that \gls{prb}. On the \gls{dl}, in contrast, the power belongs to the \gls{bs}. This difference between the \gls{ul} and \gls{dl} is key from an \gls{rrm} perspective as we shall see and this means that the numerous results and techniques developed for the \gls{dl} cannot be used as such on the \gls{ul}.
\gls{rrm} can be done on a per-\gls{prb} basis on the \gls{dl} by allocating,  a priori, the \gls{bs} power equally among the \glspl{prb} in a \gls{ts} \cite{Hussein-2024}, while on the \gls{ul}, the \gls{rrm} must be done over all \glspl{prb} in a \gls{ts} at once \cite{Haseen-2024}. This is because \gls{pa} on the \gls{ul} for a given user cannot be carried out before knowing in which \glspl{prb} (over the entire \gls{ts}) the user is selected for transmission.

An RRM process that is often ignored in the literature is \gls{mcs} selection, i.e., the process by which rate adaptation is performed by selecting an MCS out of a finite set of pre-defined MCSs based on the outcomes of \gls{ss} and power management. Instead, papers use a Shannon-based rate adaptation, which is much less realistic~\cite{Hussein-2024}.

Solving the \gls{rrm} problem for \gls{mu-mimo} systems to optimality is impractical even for a moderately sized system (e.g., for 20 users) on both the \gls{dl} \cite{Chen-2021} and the \gls{ul} \cite{Haseen-2024}, due to the need for exhaustive searches to find the best  \gls{ss}. Previous works have compared \gls{ctrone}, \gls{bd} and \gls{ctrf} on the \gls{dl} by proposing and employing offline \emph{\gls{gus}} heuristics \cite{Bjornson-2013, Boccardi-2007, Shi-2008, Joao-2025, Shen-2006}.

To the best of our knowledge, no work has studied the strategies \gls{ctrone}, \gls{bd} and \gls{ctrf} and compared their performance within the same \gls{ul} framework while taking \gls{pf} and all \gls{rrm} processes into account. We propose to adapt \gls{gus} used on the \gls{dl} to the \gls{ul} given its proven performance on the \gls{dl} \cite{Wang-2008} and wide applicability. However, such a search must be done over all the \glspl{prb} at once in a \gls{ts} (there are in general many dozens of \glspl{prb} in a \gls{ts}), which increases the search complexity greatly compared to the \gls{dl}. Another distinction between the DL and UL appears when considering fairness among users; since the \gls{rrm} procedure is done over the entire \gls{ts} at once on the UL, weights for fairness are updated every TS as opposed to every PRB on the \gls{dl}~\cite{Haseen-2024}. Many \glspl{ts} must be then executed to allow the evolution of the weights on the \gls{ul}, which means that \gls{ul} heuristics have to be designed for speed. Indeed, heuristics on the \gls{dl} had to be run over hundreds of \glspl{prb} while on the \gls{ul}, they will need to be run over thousands. 

We are now ready to state our \glspl{rq}:

\begin{enumerate}[label=\textbf{RQ \arabic*}:, leftmargin=4em]
\item How to formulate the \gls{ul} \gls{rrm} problem for the \gls{bf} strategies \gls{bd}, \gls{ctrone} and \gls{ctrf}, considering \gls{mcs}-based rate adaptation, \gls{pf} and power management over a \gls{ts}? Can this problem be solved as such?
\item If it cannot be solved as such, can we  extend, in a computationally efficient way, the \gls{gus} heuristic developed for the \gls{dl} to the \gls{ul}?
\item What are the performance gains brought by \gls{ctrf}  compared to the simpler strategies \gls{bd} and \gls{ctrone} on the \gls{ul} for different \gls{3gpp} scenarios and system parameters?  Are those gains scenario-dependent?
\item How do different power management schemes compare on the \gls{ul} of a \gls{mu-mimo} system with \gls{ma} users? 
\end{enumerate}
To answer these \glspl{rq}, we make the following contributions:

\begin{itemize}
    \item \textbf{Formulation of the \gls{ul} \gls{rrm} problem over a \gls{ts}:} this formulation applies to the  three \gls{bf} strategies \gls{bd}, \gls{ctrone} and \gls{ctrf}, considering \gls{mcs}-based rate adaptation, \gls{pf} and power management. The problem is extremely large and cannot be solved even for small but  meaningful systems and hence we need to resort to heuristics.
    \item \textbf{Efficient per-\gls{ts} \gls{gus} Design and Implementation:} We propose an efficient per-\gls{ts} \gls{gus} heuristic for the \gls{ul}, considering all key \gls{rrm} processes, that can be used for \gls{bd}, \gls{ctrone} and \gls{ctrf}. The search takes into account practical aspects, such as \gls{pf} and a \gls{mcs}-based rate function.  Furthermore, the search utilizes the faster \gls{zf} \gls{bf} computation methods developed in \cite{Wang-2008, Shi-2008} and efficient power management. 
    \item \textbf{Judicious Computation Technique:} Despite the efforts to make the search efficient by using fast \gls{zf} \gls{bf} and efficient power management, the number of rate computations at each search iteration can be very large as will be seen in Section~\ref{sec:gus}. Then, we propose a smart computation method to reduce the computational cost of \gls{gus} by judiciously reusing rates computed in past search iterations whenever possible. This aggressive reusing method can make the heuristic faster while not changing the performance (in terms of rates) since it is only smart in what to recompute. For example, for \gls{ctrone} and \gls{bd}, we show that the run time can be reduced by 11\% and 42\%, respectively, when the search is equipped with the proposed rate-reusing method. This allows us to study  the impact of  different system parameters and scenarios in a reasonable time.
    \item \textbf{Extensive Numerical Study}: Using the proposed \gls{gus} heuristic, we conduct a comprehensive evaluation of \gls{ctrf}, \gls{ctrone} and \gls{bd} over many \glspl{ts} for two different \gls{3gpp} scenarios, namely \gls{rma} and \gls{uma}, across different system parameters. We provide valuable  information that can be used for the design of real-time \gls{ul} \gls{rrm} heuristics. Particularly, we show that, while \gls{ctrf} outperforms \gls{bd} and \gls{ctrone} across all scenarios and system parameters,  \gls{ctrone} or \gls{bd} could replace \gls{ctrf} in some cases if \gls{ctrf} is deemed too complex for real-time settings. For instance, for \gls{rma}, \gls{bd} and \gls{ctrone} are alternatives to \gls{ctrf} for small and large numbers of users, respectively, given their comparable performance. For \gls{uma}, \gls{ctrone} presents a similar performance to \gls{ctrf} across all system parameters, emerging as a choice over \gls{ctrf}. We also show that the numbers of antennas at the \gls{bs} and at the users, and the power budget per user can significantly impact the performance of the \gls{zf} \gls{bf} strategies. 
    \item \textbf{Impact of Simpler Power Management Scheme:} Finally, we consider a simple   power management scheme that shares the power equally among all streams of a user over a \gls{ts} and compare its performance, for each \gls{zf} \gls{bf} strategy, with the more complex power management scheme used so far.  We show that  \gls{bd} performance is the most impacted  by the simpler scheme (decrease of 35\% in performance) relative to the other strategies (decrease of 15\% for \gls{ctrone} and \gls{ctrf}).
\end{itemize}

The paper is organized as follows. Section~\ref{sec:background} introduces the system model and presents the related work. Section~\ref{sec:rrm} describes the \gls{ul} \gls{rrm} in detail. Section~\ref{sec:gus} presents the proposed \gls{ul} \gls{gus} heuristic. Finally, Section~\ref{sec:results} contains the numerical results and discussions. Section~\ref{sec:conclusion} concludes the paper with final remarks and avenues for future research. 

\textit{Notation:} Calligraphic letters, such as \(\mathcal{A}\), denote sets, with \(|\mathcal{A}|\) indicating the cardinality of set \(\mathcal{A}\). Upper case bold letters, such as \(\mathbf{M} \in \mathbb{C}^{M \times N}\), indicate a matrix of dimension $M$ by $N$ with complex numbers. Table~\ref{tab:notation} presents the main notation used throughout this work.

\section{System Model and Related Works} \label{sec:background}

In this section, we present the system model and discuss related works and how our research fills some of the gaps in the literature.
  
\subsection{System Model} \label{subsec:system}
We consider the \gls{ul} of a single \gls{mu-mimo} \gls{ofdma}-based cell. \gls{ofdma} divides the system bandwidth into subchannels and the time into \glspl{ts}. The bandwidth is divided into $C$ subchannels, each with a bandwidth of $B_C$, where $\mathcal{C} = \{1,\dots,C\}$ is the set of subchannels. A \gls{prb} consists of a subchannel $c \in \mathcal{C}$ in a TS $t$.  The \gls{bs} and users have $M_B$ and $M_U$ antennas, respectively\footnote{We assume an equal number of antennas per user for simplicity, but this study can be extended to a non-equal setting.}. Under \gls{zf} \gls{bf} with \gls{ma} users, each user can transmit up to $M_U$ independent data streams in a given \gls{prb}, while the \gls{bs} can receive up to $M_B$ streams in a \gls{prb} \cite{Guthy-2009}. The users transmit at a carrier frequency $f_c$ with a per-\gls{ts} power budget of $P_U$. We assume no signaling overhead, error-free \gls{csi}, and full buffers at the users so that data is always available for transmission. The channel model is described in Section~\ref{sec:results}. Table~\ref{tab:notation} summarizes the notations used throughout this work.

\subsection{Related Works}

\begin{table}[t]
    \centering
    \caption{Main notations: subscript $u,s$ means stream $s$ of user $u$; superscript $c$ (resp. $t$) refers to PRB $c$ (resp. TS $t$)} 
    \begin{tabular}{|c|c|}
        \hline
         Notation & Description  \\
         \hline
         $M_B$ and $M_U$& Number of antennas at the \gls{bs} and at the users \\ \hline
         $\mathcal{U}$,  $\mathcal{M}_U$ and $\mathcal{C}$ & Set of users, stream indices and subchannels in a TS \\ \hline
         $v^c_{u,s}$ & Binary variable for stream selection \\ \hline
        $\mathbf{V}^c$; $\mathcal{V}$ & Allocation matrix; Allocation set over the TS\\ \hline
         $\mathbf{H}_u^c$; $\mathcal{H}$ & Channel matrix; Set of channel matrices over the TS \\ \hline
         $w^t_u$ and $R^t_u$ & Weight and moving average per-TS rate for a TS \\ \hline
         $r^t_u$ & Rate transmitted in a TS\\ \hline
         $W$ & Window of TSs for fairness  \\ \hline
         $P_U$ & Per-TS power budget per user\\ \hline
         $P^c_{u,s}, r^c_{u,s}, E^c_{u,s}$ & Power, rate and Effective channel \\ \hline
         $L$; $\Gamma_l$ and $\zeta_l$ & Number of MCSs; SNR and rate for MCS $l$ \\ \hline
         $\tau^c_{u,s}$ & Power for the highest MCS level \\ \hline
         $f(\cdot)$ & Function mapping SNR to rate \\ \hline
         $A$ and $D$ & Coefficients of the fitted MCS-based rate function \\ \hline
    \end{tabular}
    \label{tab:notation}
\end{table}

\begin{table*}[ht]
    \centering
    \caption{Synthesis of related works. Notation: Shannon-based (SB), User Antenna Setting (UAS), Sum-Power Minimization (SPM).}
    \begin{tabular}{c|c|c|c|c|c}
         Work &  Rate Adaptation & Subchannels & Criterion & BF Strategy & UAS \\
         \hline
         \cite{Haseen-2024} &  MCS & \textbf{Multiple} & PF & \gls{ctrone}=\gls{bd} & SA\\
         \cite{Guthy-2009} &   SB & Single & WSR & Similar to \gls{ctrf} & \textbf{MA} \\
         \textbf{\cite{Andrew-2022}} & MCS & \textbf{Multiple} & Minimum Rate & BD & \textbf{MA} \\
         \cite{Li-2007} &  SB & Single & SPM & MRT/SIC & \textbf{MA}\\
        \cite{Feres-2023}  & SB & \textbf{Multiple} & SR & MMSE & SA\\
         \textbf{\cite{Zhang-2024}}  & SB & \textbf{Multiple} & PF & MMSE/MRT & \textbf{MA} \\
         \cite{Meng-2018}  & SB & \textbf{Multiple} & PF & MMSE & SA\\
         \textbf{\cite{Zhang-2005}}  & MCS & \textbf{Multiple} & SPM/QoS & MRT/MMSE/ZF & \textbf{MA}\\
         Ours & MCS & \textbf{Multiple} & PF & \gls{ctrone}, \gls{bd}, \gls{ctrf} & \textbf{MA} \\ 
         
         \hline
    \end{tabular}
    \label{tab:ULpapers}
\end{table*}

As discussed previously, the \gls{ul} \gls{rrm} procedure should be done over all \glspl{prb} of a \gls{ts} at once due to the power management, which increases the difficulty when compared to the \gls{dl} (which can be done on a \gls{prb} basis). In the following, we discuss works on the \gls{ul} of \gls{mu-mimo} systems. 

The most relevant works to ours were conducted in \cite{Haseen-2024}, \cite{Guthy-2009} and \cite{Andrew-2022}. In \cite{Haseen-2024}, an offline \gls{gds} heuristic over the \gls{ts} was proposed for \gls{sa} users with \gls{zf} \gls{bf}. \gls{gds} can be readily used for \gls{ctrone} as \gls{ma} users with one stream can be interpreted as \gls{sa} users, but its extension to \gls{bd} or \gls{ctrf} is not clear. Also, \gls{gds} has a constraint on the number of users in the cell which must not be greater than the number of \gls{bs} antennas. A \gls{gus} heuristic based on ZF with multiple streams per user, where the \gls{bf} matrices were optimized during the search, was proposed in \cite{Guthy-2009}. However, it was based on a single \gls{prb}, ignoring completely the major issue of power management on the \gls{ul}, and its design was based on the Shannon rate function. \cite{Andrew-2022} considered a multi-channel \gls{mu-mimo} system employing \gls{bd} with \gls{ma} users, focusing on maximizing the minimum rate offered to all users. However, the work considered a very simple heuristic where user selection is based on allocating  the same number of subchannels to each user, and power management is based on equal \gls{pa}. 

The \gls{ul} was also addressed in other significant works, which are less relevant to our study but are reported for completeness. For instance, in \cite{Li-2007}, the sum-power minimization problem with rate constraints was investigated with \gls{mrt} \gls{bf} for \gls{ma} users (i.e., each user utilized only its strongest stream) and \gls{sic} at the \gls{bs}. \cite{Feres-2023} proposed an unsupervised learning user selection method with \gls{mmse} receivers at the \gls{bs}, but it did not consider fairness and was designed for \gls{sa} users. In \cite{Zhang-2024}, the scheduling \gls{ul} problem over multiple \glspl{prb} in a \gls{ts} was addressed with \gls{mrt} \gls{bf} for \gls{ma} users and \gls{mmse} receivers at the \gls{bs}, but the solution design was based on the Shannon rate function and single-stream users. In \cite{Meng-2018}, a grouping-based user scheduling method for \gls{pf} and \gls{sa} users with \gls{mmse}-based decoding at the \gls{bs} was proposed. \cite{Zhang-2005} examined the minimization of overall transmit power over multiple \glspl{prb} while guaranteeing the fulfillment of each user's \gls{qos} requirements with \gls{mrt} \gls{bf} for \gls{ma} users and different receivers, such as \gls{mmse} and \gls{zf}.

Table~\ref{tab:ULpapers} summarizes the main characteristics of the papers discussed above. From this discussion, it is clear that no work has evaluated the performance of the \gls{zf} \gls{bf} strategies on the \gls{ul} of a multi-channel \gls{mu-mimo} system with \gls{ma} users and considering all key \gls{rrm} processes carefully.  Our work intends to answer the research questions above. 

\section{Uplink Radio Resource Management} \label{sec:rrm}

In this section, we introduce the UL RRM over an arbitrary TS $t$, omitting index $t$ and referring to a subchannel as a \gls{prb} for ease of notation. Specifically, we discuss each \gls{rrm} process independently from  \gls{ss}, \gls{zf} \gls{bf}, power management to \gls{mcs} selection. Also, we formulate and discuss the general \emph{joint} \gls{ul} \gls{rrm} problem over the \gls{ts}.  
We explain why this joint problem cannot be solved for realistic systems and then propose heuristics to obtain good feasible solutions using the problem structure.

We start by presenting some important notation. Let $\mathcal{U}$ be the set of users. Let $\mathbf{H}^c_u \in \mathbb{C}^{M_U \times M_B}$ be the channel matrix for user $u\in \mathcal{U}$ in \gls{prb} $c \in \mathcal{C}$ and $\mathcal{H} = \{\mathbf{H}_u^c\}_{u \in \mathcal{U}, c \in \mathcal{C}}$ be the channel matrices set for all users and \glspl{prb} in the TS. 

\subsection{Stream Selection} The \gls{ss} process selects which users and streams will be allowed to transmit in each \gls{prb}. On the UL, \gls{ss} has to be done over all \glspl{prb} at once as opposed to sequentially one \gls{prb} at a time as in the \gls{dl}, because power management, which will be discussed below, depends on which \glspl{prb} are allocated to a user \cite{Hussein-2024, Haseen-2024}. The stream selection is reduced to a user selection problem for \gls{ctrone} and \gls{bd}, since for each scheduled user $u\in \mathcal{U}$ in a \gls{prb} $c \in \mathcal{C}$, the \gls{bs} enables only the strongest stream (resp. all $M_U$ streams) for \gls{ctrone} (resp. \gls{bd}). Under \gls{ctrf}, any combination of streams can be enabled for each user in any \gls{prb} (up to $M_U$ streams per user per \gls{prb}).

Let the stream indices take any integer value from $1$ to $M_U$ with stream $1$ being the strongest and stream $M_U$ the weakest (see \cite{Shi-2008, Boccardi-2007} for details on stream strength) and let $\mathcal{M}_U = \{1,\dots,M_U\}$ be the stream indices set. Let $v^c_{u,s} \in \{0,1\}$ be the variable indicating stream selection, where $v^c_{u,s} = 1$  if stream $s \in \mathcal{M}_U$ of user $u \in \mathcal{U}$ is selected in PRB $c \in \mathcal{C}$ and $v^c_{u,s} = 0$ otherwise. For any \gls{zf} \gls{bf} strategy, another restriction is on the number of scheduled streams per \gls{prb}, which must be no greater than $M_B$. Then, in a given PRB $c \in \mathcal{C}$, we have the following constraints:
\begin{itemize}
    \item For \gls{ctrone}, $v^c_{u,s} = 0, \forall u \in \mathcal{U}$ and  $1<s\leq M_U$;
    \item For \gls{bd}, $\sum_{s \in \mathcal{M}_U} v^c_{u,s}= v^c_{u,1}M_U, \forall u \in \mathcal{U}$;
    \item For \gls{ctrf}, $\sum_{s \in \mathcal{M}_U} v^c_{u,s} \leq M_U, \forall u \in \mathcal{U}$;
    \item For \gls{ctrone}, \gls{bd} and \gls{ctrf}, $\sum_{u \in \mathcal{U}} \sum_{s \in \mathcal{M}_U}v^c_{u,s} \leq M_B$.
\end{itemize}   
We define the matrix $\mathbf{V}^c = (v^c_{u,s})$,  $\forall c \in \mathcal{C}, \forall u\in \mathcal{U}, \forall s \in \mathcal{M}_U$, and $\mathcal{V}$ to be the allocation set $\mathcal{V} = \{\mathbf{V}^c\}_{c \in \mathcal{C}}$ {over the whole TS. In the following, if $v^c_{u,s} = 1$, we will say that stream $s \in \mathcal{M}_U$ of user $u \in \mathcal{U}$ has been selected (equivalently scheduled) in \gls{prb} $c \in \mathcal{C}$. Likewise, we will say that a user $u$  is selected (scheduled) in \gls{prb} $c$ if at least one of its streams is selected for transmission (i.e., $\exists s \in \mathcal{M}_U$ such that $v^c_{u,s}=1$).

\subsection{Zero-Forcing Beamforming} \gls{zf} \gls{bf} enables the transmission of interference-free data streams\footnote{Under error-free \gls{csi}, \gls{zf} \gls{bf} nullifies interference.} from multiple users in the same \gls{prb} (at most $M_B$ streams) on the \gls{ul}, where, with \gls{ma} users, each user may have multiple streams (at most $M_U$) \cite{Guthy-2009}. Inter-user interference is nullified to the detriment of the co-scheduled users' effective channels. The more correlated the selected users in the same \gls{prb} are, the more their effective channels are reduced to achieve interference-free transmission, hence the need for appropriate \gls{ss} to schedule users with low mutual correlation \cite{Chen-2021, Yi-2011}. 

Given an allocation set $\mathcal{V}$, ZF BF is performed for each PRB $c \in \mathcal{C}$ independently (and hence can be done in parallel). For a given PRB, it is performed jointly over the selected streams to obtain the effective channels $E^c_{u,s}(\mathbf{V}^c, \{\mathbf{H}^c_k\}_{k \in \mathcal{U}}), \forall u \in \mathcal{U}, \forall s \in \mathcal{M}_U$. We use the convention that $E^c_{u,s}(\mathbf{V}^c, \{\mathbf{H}^c_k\}_{k \in \mathcal{U}}) = 0$ if a stream is not selected, i.e., if $v^c_{u,s} = 0$. The effective channels depend on which streams are selected and on the channel matrices of the users scheduled for transmission in the \gls{prb}. For brevity, we will write the effective channels as $E^c_{u,s}(\mathbf{V}^c)$ in the following.

Unfortunately, computing the effective channels for a given PRB, given a set of scheduled streams, is cumbersome. It requires singular value decompositions of matrices, which makes it difficult to incorporate as such in an optimization problem. Several studies have proposed numerical algorithms to efficiently compute the effective channels under \gls{zf} \gls{bf} \cite{Spencer-2004, Boccardi-2007, Shi-2008, Wang-2008, Tran-2012-2}. In this work, we employ the approach developed in \cite{Shi-2008, Wang-2008} in our \gls{gus} heuristic in Section~\ref{sec:gus}. This method fits the iterative nature of our heuristic since it reuses \gls{bf} computations from previous iterations to speed up the \gls{zf} \gls{bf} computations in the current iteration, as will be discussed in Section~\ref{sec:gus}.  Also, the method can be used for \gls{ctrone}, \gls{ctrf} and \gls{bd}. Other ways of computing the effective channels can be found in \cite{Boccardi-2007, Spencer-2004, Tran-2012-2}.

\subsection{Power Management} 
Given an allocation set $\mathcal{V}$ and a BF strategy, we can compute the effective channels $E^c_{u,s} (\mathbf{V}^c)$ for all streams $s \in \mathcal{M}_U$ of all users $u \in \mathcal{U}$ in each \gls{prb} $c \in \mathcal{C}$ and then execute power management over the \gls{ts} to obtain the allocated power $P^c_{u,s}$ for stream $s$ of user $u$ in \gls{prb} $c$, with the convention that $P^c_{u,s} = 0$ if $v^c_{u,s} = 0$. Then, since the product between $P^c_{u,s}$ and $E^c_{u,s} (\mathbf{V}^c)$ is the \gls{snr} for stream $s$ of user $u$ in \gls{prb} $c$,   a rate function $f(.)$ can be applied to the SNR to obtain the  rate $r^c_{u,s}$. Note that there is no intra-user interference due to \gls{zf} \gls{bf}. We now present the rate function.

\subsection{Rate Function and MCS Selection}
In most papers, the rate function is chosen to be based on the Shannon formula \cite{Guthy-2009, Zhang-2024}. However, in a practical system, there is a set of available \glspl{mcs} for selection. Each \gls{mcs} requires a minimum \gls{snr} to achieve a fixed rate for a target \gls{bler} \cite{Feminias-2016}. MCS selection is about choosing the highest \gls{mcs} for each selected stream over a \gls{prb} given its \gls{snr}. Let $L$ be the number of available \glspl{mcs}, let $\zeta_l$ and $\Gamma_l$ be the achievable rate and required \gls{snr} for \gls{mcs} $l \in \{1,\dots,L\}$, where $\zeta_l$ and $\Gamma_l$ are increasing sequences on $l$, then the \gls{mcs}-based rate function $f(\cdot)$ can be written as
    \[ f(SNR) = \begin{cases} 
          0 & SNR < \Gamma_1; \\
          \zeta_l & \Gamma_l \leq SNR < \Gamma_{l+1}, \; l \in \{1, L-1\};\\
          \zeta_L & SNR \geq 
          \Gamma_L. 
       \end{cases}
    \]
Note that $f(\cdot)$ is a piecewise constant function, and is neither concave nor differentiable.

In the following, we formulate the joint \gls{ul} \gls{rrm} problem over a \gls{ts} for our system. This problem is more general than the one in \cite{Haseen-2024} because \gls{ma} users are considered instead of \gls{sa} users. Considering  \gls{ma} users makes the problem more complex since each user can have multiple streams (with \gls{bd} and \gls{ctrf}) and in this case,  power management includes not only \gls{pa} but also \gls{pd} (while only \gls{pa} is needed for \gls{sa} users).

\subsection{Problem Formulation}
We are now ready to formulate the joint RRM problem on the \gls{ul}, for a given \gls{ts}, where fairness is taken into account by using weights $w_u$'s that reflect how much rate has been received by each user $u$ in a window of size $W$~TSs (please see Section~IV in  \cite{Haseen-2024} for more details). These weights are computed from previous \glspl{ts} results as will be discussed in Section~\ref{sec:results}. Let $\mathcal{W} = \{w_u\}_{u \in \mathcal{U}}$ be the set of all weights. 
Given $\mathcal{H}$ and $\mathcal{W}$ as well as a sufficiently large positive constant $A$, under \gls{zf} \gls{bf}, the \gls{ul} \gls{rrm} over a \gls{ts} can be written as the following \gls{wsr} maximization problem:
\begin{align}
    & \underset{(P^c_{u,s}), (v_{u,s}^c), (r^c_{u,s}), (E^c_{u,s}(\mathbf{V}^c))}{\max} \sum_{u \in \mathcal{U}}\sum_{c \in \mathcal{C}}\sum_{s \in \mathcal{M}_U} w_u r^{c}_{u,s} \label{eq:wsr} \\
    &  \sum_{c \in \mathcal{C}}  \sum_{s \in \mathcal{M}_U} P^{c}_{u,s} \leq P_{U}, \;\; \forall u  \label{eq:sumpower} \\
    & P^{c}_{u,s} \geq 0, \;\; \forall u, c, s  
    \label{eq:nonneg} \\
    & r^c_{u,s} = f\left(P^c_{u,s} E^c_{u,s} (\mathbf{V}^c)\right) \;\; \forall u, c, s  \label{eq:rate} \\
    & E^c_{u,s}(\mathbf{V}^c) \leq v^c_{u,s} A, \;\; \forall u, c, s  \label{eq:effec} \\
    & v^c_{u,s} \in \{0,1\}, \;\; \forall u, c, s  \label{eq:bin} \\
    & \sum_{u \in \mathcal{U}} \sum_{s \in \mathcal{M}_U}v^c_{u,s} \leq M_B, \;\; \forall c  \label{eq:zfbf} \\
    & \text{\gls{ctrone}}: \; v^c_{u,s} = 0, \;\; \forall u, c, s \neq 1  \label{eq:ctrone}\\
    & \text{\gls{bd}}: \; \sum_{s \in \mathcal{M}_U} v^c_{u,s} = 0 \; \text{or} \; \sum_{s \in \mathcal{M}_U} v^c_{u,s}= M_U, \;\;\forall u,c  \label{eq:bd} \\
    & \text{\gls{ctrf}}: \; \sum_{s \in \mathcal{M}_U} v^c_{u,s} \leq M_U, \;\; \forall u,c \label{eq:ctrf}
\end{align}
Recall that $\mathbf{V}^c$ is the matrix  $(v^c_{u,s})$ of dimension $|\mathcal{U}| \times M_U$ representing the stream selection for \gls{prb} $c$. The variables $v^c_{u,s}$'s are binary, which makes the problem an integer one.  Constraint~\eqref{eq:sumpower} is the sum-power constraint per user over the \gls{ts}, coupling the \glspl{prb} in the \gls{ts} for each user. This constraint is the reason why the \gls{ul} \gls{rrm} has to be done over the entire \gls{ts} as opposed to per \gls{prb} on the \gls{dl}. 
Equation~\eqref{eq:rate} maps the \gls{snr} of a stream to a rate with the rate function $f(\cdot)$. Recall that the \gls{mcs}-based function $f(\cdot)$ is not concave, which makes the problem non-convex. Implicit in this formulation are the computations of the effective channels, which, as discussed earlier, cannot be expressed in a simple algebraic form. This also makes the problem non-convex. Constraint~\eqref{eq:effec} guarantees that $E^c_{u,j}(\mathbf{V}^c)=0$ if stream $s$ of user $u$ is not selected in PRB $c$. 
Constraint~\eqref{eq:zfbf} originates from the \gls{zf} \gls{bf} restriction on the number of scheduled streams per \gls{prb}, which must be no greater than $M_B$. Constraint~\eqref{eq:ctrone} is only used for \gls{ctrone} and ensures that only the strongest stream of each selected user is enabled, while Constraint~\eqref{eq:bd} is introduced for \gls{bd}, guaranteeing that all the streams are enabled for each selected user. For \gls{ctrf}, Constraint~\eqref{eq:ctrf} limits the sum of $v^c_{u,s}$ for each user and \gls{prb} to be less than or equal to $M_U$. 

This problem is a large integer and highly non-convex problem that needs computations of effective channels that can only be done using a numerical approach. In the next section, we discuss why obtaining feasible solutions for such a problem is impractical even for small systems. 

\subsection{Why a straightforward solution is not applicable} \label{sec:originalproblem}
We note that, given a stream selection, i.e., a matrix  $\mathbf{V}^c$ for all $c \in \mathcal{C}$, then the effective channels can be computed on each \gls{prb}, and then the problem becomes a set of $|\mathcal{U}|$ much simpler \textbf{per-user} power management (joint power allocation and power distribution) problem (JPM in short), i.e., 
\begin{align}
    &\text{\textbf{JPM}} (u, \{E^c_{u,s} (\mathbf{V}^c)\}_{c, s}): \label{eq:jpapd}
     \underset{P^c_{u,s}, r^c_{u,s}}{\max} \sum_{c \in \mathcal{C}}\sum_{s \in \mathcal{M}_U} r^{c}_{u,s} \\ 
    & \text { s.t. } \sum_{c \in \mathcal{C}}  \sum_{s \in \mathcal{M}_U} P^{c}_{u,s} \leq P_{U} \label{eq:powerconst1} \\
    & P^{c}_{u,s} \geq 0, \;\; \forall c,s 
    \label{eq:powerconst2} \\
    & r^c_{u,s} = f\left(P^c_{u,s} \times E^c_{u,s} (\mathbf{V}^c)\right), \;\; \forall c,s \label{eq:rateconst1}
\end{align}
JPM is no longer an integer problem. It is non-convex because of the \gls{mcs}-based rate function $f(.)$. Nevertheless, it can easily be convexified as done in \cite{Andrew-2022, Yuan-2024} by fitting the rate function with a concave function (either an upper-bound or a tight approximation).

Hence, solving the original problem could in theory be done by doing an exhaustive search on the set of matrices $\{\mathbf{V}^c\}_{c \in \mathcal{C}}$ and by computing, for each such set, the effective channels for each stream of each user in each PRB and then solving the convexified version of the per-user JPM problem to finally compute the WSR over the TS. Comparing the upper-bound with the feasible solution obtained with a tight approximation of the rate function would allow us to validate the goodness of the feasible solution. Clearly, this is too time-consuming, especially since the computations of the effective channels are cumbersome and exhaustive searches on $\{\mathbf{V}^c\}_{c \in \mathcal{C}}$ are impractical even for small systems since there are too many different stream selections over the multiple \glspl{prb} of a \gls{ts}. Instead of an exhaustive search, we will do a directed search, which will still be cumbersome but will be tractable and will allow us to obtain a feasible solution to the problem.

\section{Uplink Greedy-Up Search Heuristic} \label{sec:gus}

We present in this section our proposed offline per-\gls{ts} greedy up heuristic, \gls{gus},  that considers all key \gls{rrm} processes and can be used for \gls{ctrone}, \gls{bd} and \gls{ctrf}. The heuristic incorporates realistic considerations, i.e., an \gls{mcs}-based rate function and \gls{pf}. It operates in an iterative greedy-up fashion to build a feasible solution over the \gls{ts} including a selection of streams per \gls{prb}, along with their allocated power and rate. In each search iteration, for each PRB, we try all streams (resp. all users) one by one among those not yet selected in that \gls{prb} for \gls{ctrf} and \gls{ctrone} (resp. for \gls{bd}). At the end of the iteration, the pair among all the stream (resp. user) and \gls{prb} pairs that yields the largest \gls{wsr} over the \gls{ts} when combined with the already selected streams (resp. users) in all \glspl{prb} is added to the set of scheduled pairs of streams (resp. users) and \glspl{prb}. Note that for \gls{ctrone}, only the strongest stream of each user can be selected. 



Efficiency is at the core of the heuristic design and implementation, as it must be tractable over many \glspl{ts} to allow the weights to evolve for performance evaluations under fairness. To illustrate that, consider \gls{ctrf} in a system with 40 users, each with 4 antennas, and 78 \glspl{prb} in a \gls{ts}. Consider a search iteration in which a total of 30 streams have already been selected across the 78 \glspl{prb}. Then, to select just one new stream in a \gls{prb}, we have to assess  130 streams (40 × 4 streams in total minus 30 already selected streams) in 78 \glspl{prb}, one at a time (more than $10^3$ assessments), and choose the stream that provides the highest \gls{wsr} over the \gls{ts}. In each of those assessments, possibly one \gls{zf} \gls{bf} computation and several power computations (one per user affected by the assessment) have to be carried out, as will be discussed shortly. Therefore, to reduce its computational cost, we have equipped \gls{gus} with efficient \gls{zf} \gls{bf} and power computation techniques as well as a novel iterative rate-reusing method.

\subsection{Heuristic Design and Implementation} \label{sec:search} 
We use \gls{ctrf} in the following to present \gls{gus}, since it is the most intricate case. We will discuss \gls{gus} for BD and \gls{ctrone} in Section~\ref{sec:toto}. In the case of \gls{ctrf}, the heuristic executes an unconstrained greedy-up search on the stream sets per \gls{prb} over the \gls{ts}. The heuristic starts by initializing $v^b_{k,j} \leftarrow 0, \forall k \in \mathcal{U}, \forall b \in \mathcal{C}, \forall j \in \mathcal{M}_U$, meaning that, initially, there is no scheduled stream. Let $WSR_{max}$ be the largest \gls{wsr} over the \gls{ts} registered so far during the search, and $\hat{v}^b_{k,j}$ be the best stream allocation over the \gls{ts} yielding $WSR_{max}$. They  are initialized as $WSR_{max} \leftarrow 0$ and $\hat{v}^b_{k,j} \leftarrow 0, \forall k \in \mathcal{U}, \forall b \in \mathcal{C}, \forall j \in \mathcal{M}_U$. At each search iteration, we add one stream in one of the \glspl{prb}. 
Specifically, at the beginning of a search iteration, we assess each candidate for selection $(u,c,s)$ where $s$ is a   stream  of  user $u$ in \gls{prb} $c \in \mathcal{C}$ such that $v^c_{u,s} = 0$ (i.e., $s$ has not yet been scheduled in $c$).  Clearly, the number of such candidates for selection can be very large, especially at the beginning of the search. The selected candidate for scheduling is the one that yields the largest \gls{wsr} over the \gls{ts} when scheduled together with the already selected streams over all \glspl{prb}. 

Let $WSR^c_{u,s}$ be the \gls{wsr} over the \gls{ts} when candidate $(u,c,s)$ is assessed. For the sake of computing $WSR^c_{u,s}$, assume stream $s$ of user $u $ has been selected temporarily in \gls{prb} $c$ (i.e., $v^c_{u,s} \leftarrow 1$). 

To compute $WSR^c_{u,s}$ for candidate $(u,c,s)$, the steps are:
\begin{itemize}
    \item \textbf{\gls{zf} \gls{bf} Computation:}
    To improve the search computational efficiency, we reuse effective channel values (when possible) and employ fast \gls{zf} \gls{bf} computation methods \cite{Wang-2008, Shi-2008}. 
    \item \textbf{Rate Computation:} To obtain the rates for each selected stream over the \gls{ts}, given the addition of candidate $(u,c,s)$, we need to execute power management (i.e., \gls{pa} and \gls{pd}) for each user $k \in \mathcal{U}$ that has been selected in \gls{prb} $c$ (i.e., $\forall k \in \mathcal{U}~\text{s.t.}~\exists j \in \mathcal{M_U}~\text{s.t.}~ v^c_{k,j} = 1$) and obtain the power allocated for each of its selected stream $P^b_{k,s}, \forall j \in \mathcal{M}_U, \forall b \in \mathcal{C}$ over all channels $c$ of the \gls{ts}. 
Executing power management for each user $k \in \mathcal{U}$ selected in \gls{prb} $c$ is necessary because the effective channel values for the currently selected streams in \gls{prb} $c$ most likely change when adding stream $s$ of user $u$.

To execute power management, we could solve JPM defined in Section~\ref{sec:originalproblem} for each of those users. However, due to the nature of the \gls{mcs}-based function $f (\cdot)$, the problem is non-convex, preventing the use of efficient algorithms such as \gls{wf} \cite{Xing-2020}. Instead, we follow a similar approach to \cite{Andrew-2022, Yuan-2024} and fit $f(\cdot)$ with a smooth concave function, specifically $A\log(1 + D\gamma)$, where $A$ and $D$ are non-negative coefficients, where $\gamma$ is the \gls{snr}. Using the fitted function,  JPM over the \gls{ts} can be rewritten as, for a user $k \in \mathcal{U}$,
\begin{align}
    & \overline{\text{\textbf{JPM}}}(k, \{E^b_{k,j} (\mathbf{V}^b)\}_{b, j}): \nonumber\\
    & \underset{P^b_{k,j}}{\max} \sum_{b \in \mathcal{C}}\sum_{j \in \mathcal{M}_U} A\log(1 + DE_{k,j}^{b} (\mathbf{V}^b)P^{b}_{k,j}) \\
    & \text { s.t. } \sum_{b \in \mathcal{C}}\sum_{j \in \mathcal{M}_U} P^{b}_{k,j} \leq P_{U} \label{eq:fitconst1}; \\
    & \; 0 \leq P^{b}_{k,j} \leq \tau^{b}_{k,j}, \forall b \in \mathcal{C}, \forall j \in \mathcal{M}_U, 
    \label{eq:fitconst2}
\end{align}
which can be solved efficiently by \gls{wf} using Algorithm~6 from \cite{Xing-2020}. The power is upper bounded in Constraint~\eqref{eq:fitconst2} to avoid power waste since the \gls{mcs}-based rate function is upper bounded by $\zeta_L$ for any \gls{snr} greater or equal to $\Gamma_L$ (the required \gls{snr} for $L$, the highest order \gls{mcs}). 
Specifically, $\tau^{b}_{k,j}$ is the power necessary for stream $j$ to achieve an \gls{snr} equal to $\Gamma_L$, since allowing  a larger \gls{snr} would be wasteful.

For each user $k \in \mathcal{U}$ that has a stream  selected in \gls{prb} $c$, we compute the rates $r^b_{k,j}$ using the fitted \gls{mcs}-based rate function. Note that the rate computations have to be done very often and hence we are only using some of the characteristics of the \gls{mcs}-based rate function at this stage. We will use it fully at a later stage.
\item \textbf{Computation of $WSR^c_{u,s}$}: Finally, the \gls{wsr} when adding candidate $(u,c,s)$ is  $$WSR^c_{u,s} = \sum_{k \in \mathcal{U}}\sum_{b \in \mathcal{C}}\sum_{j \in \mathcal{M}_{U}} w_k r^{b}_{k,j},$$ where we reuse the rates $r^{b}_{k,j}$ for users $k \in \mathcal{U}$ that have no streams selected in \gls{prb} $c$,  computed in previous search iterations (the rates of users that have not been selected in \gls{prb} $c$ are not affected by scheduling a new stream in that \gls{prb}). Note that we do not execute an exact \gls{mcs} selection during the search to decrease its computational cost. Although \gls{mcs} selection generally requires a small binary search, executing it several times slows the search down.

It is important to highlight the computational burden caused by this step, as many instances  of $\overline{\text{JPM}}$ must be executed for each candidate assessment.
\end{itemize}

After assessing all unselected streams across the users and \glspl{prb} in a given search iteration, we can determine the candidate $(u^*,c^*,s^*)$ that, when selected together with the already selected streams in all \glspl{prb}, yields the largest \gls{wsr}, i.e., $(u^*,c^*,s^*) = \text{argmax}_{(u \in \mathcal{U}, c \in \mathcal{C}, s \in \mathcal{M}_U)} WSR^c_{u,s}$. Note that the weights $w_u$ are used to compute the weighted sum rates. If $WSR^{c^*}_{u^*,s^*} > WSR_{max}$, then  the candidate $(u^*,c^*,s^*)$ is definitely selected (i.e., $v^{c^*}_{u^*,s^*} \leftarrow 1$) and we update $WSR_{max} \leftarrow WSR^{c^*}_{u^*,s^*}$ and $\hat{v}^{c^*}_{u^*,s^*} \leftarrow 1$, and go to the next search iteration. However, if $WSR^{c^*}_{u^*,s^*} < WSR_{max}$ but $WSR_{max}/WSR^{c^*}_{u^*,s^*} \leq \beta$ (where $\beta$ is a threshold that we choose by trial and error), we still select the candidate $(u^*,c^*,s^*)$ and continue the search (this is because the \gls{wsr} may start to increase again at some future search iteration), but we do not update $WSR_{max}$ and $\hat{v}^{c^*}_{u^*,s^*}$. Finally, if $WSR_{max}/WSR^{c^*}_{u^*,s^*} > \beta$, we stop the search. Two other evident stopping criteria for the search are when $M_B$ streams have been selected in all \glspl{prb}, i.e., $\sum_{k \in \mathcal{U}} \sum_{j \in \mathcal{M}_U}v^b_{k,j} = M_B, \;\; \forall b \in \mathcal{C}$, or when all streams of all users have been selected in all \glspl{prb}, i.e., $\sum_{j \in \mathcal{M}_U} v^b_{k,j} = M_U, \forall k \in \mathcal{U}, \forall b \in \mathcal{C}$. Upon stopping, the search returns the selected streams over the \gls{ts} corresponding to the largest $\gls{wsr}_{max}$ observed during the search (given by $\hat{v}^b_{k,j}$), along with their effective channels.

Next, we perform a last round of power management with a focus on the exact MCS-based rate function. Specifically, we execute \gls{pa} and \gls{pd} for each user $k \in \mathcal{U}$ individually over its selected streams over the \gls{ts} using first \gls{wf} (by solving $\overline{\text{JPM}}$) and using the solution as an input to  the greedy \gls{mcs}-aware heuristic from Algorithm~1 in \cite{Hussein-2024}. This \gls{mcs}-aware heuristic can improve the rates for each user by selecting better \glspl{mcs} for its scheduled streams over the \gls{ts}. We call this two-step joint \gls{pa} and \gls{pd}, \gls{tpm}. It  is computationally efficient (due to \gls{wf}) while considering the \gls{mcs} mapping (through the \gls{mcs}-aware heuristic) for performance enhancement.  With this \gls{mcs} selection, some streams in some \glspl{prb} may receive a zero rate if their \glspl{snr} are below $\Gamma_1$. Unselecting zero-rate streams in  PRBs can improve performance since the effective channels of the remaining streams may increase \cite{Hussein-2024}. We do remove all zero-rate streams at once, redo the \gls{zf} \gls{bf} computations in the \glspl{prb} where streams were removed, and redo \gls{tpm} for the users. Then, we obtain the final rate seen by user $k \in \mathcal{U}$ in the \gls{ts}, $r_k = \sum_{b \in \mathcal{C}} \sum_{j \in \mathcal{M}_U} r^b_{k,j}$. This ends the \gls{gus} heuristic for \gls{ctrf}. 

\subsection{Speeding Up the Computations: Rate-Reusing Method} \label{sec:reusing}
To compute  $WSR^c_{u,s}$ for a candidate $(u,c,s)$, we execute \gls{zf} \gls{bf} computations using a fast computation method, then the rate computations are done with \gls{wf}-based \glspl{pa} and \glspl{pd} (i.e., by solving $\overline{\text{JPM}}$) using a smooth fitted rate function (no search is necessary on the \glspl{mcs}). Nonetheless, the number of candidates in a particular search iteration can be very large, as discussed previously. We saw that for each candidate assessment, possibly one \gls{zf} \gls{bf} computation is needed, whereas several rate computations (where \gls{pa} and \gls{pd} are the sources of complexity) are required. The high computational cost from rate computations can hinder performance evaluations on the \gls{ul} for large systems over many \glspl{ts}. To reduce the number of rate computations in each search iteration and make the search more efficient, we have established rules  to decide precisely which rates must be recomputed when assessing the candidates in a given search iteration and which rates can be reused from previous search iterations.  These rules are given in Appendix~\ref{append} for \gls{bd}. They can easily be extended to \gls{ctrone} and \gls{ctrf}. They can reduce the heuristic run time by about 42\%, as shown in Section~\ref{sec:results}, while not affecting the performance as we are only being judicious in what to recompute.

\subsection{\gls{gus} for \gls{ctrone} and \gls{bd}}\label{sec:toto} 
\gls{gus} for both \gls{ctrone} and \gls{bd} is similar but simpler, with the key difference being that the search scope is over users instead of streams: for \gls{bd}, we use all $M_U$ streams when assessing and selecting a user in a particular \gls{prb}, whereas, for \gls{ctrone}, we assess and select only the strongest stream of a user in a certain \gls{prb}. Put differently, the candidates in a given search iteration for both \gls{bd} and \gls{ctrone} are not defined by specific streams; a candidate for \gls{bd} and \gls{ctrone} is ($u,c$) defined by an unselected user $u \in \mathcal{U}$ in \gls{prb} $c \in \mathcal{C}$. 
\section{Numerical Results} \label{sec:results}

The numerical results are presented in this section. We start by defining the experimental setup, presenting how the weights for \gls{pf} are updated between \glspl{ts} and the procedure used to execute many \glspl{ts}, which is necessary to allow the weights to evolve.
\subsection{Experimental Setup} \label{subsec:experimental}
We consider the \gls{3gpp}-based \gls{rma} and \gls{uma} scenarios. Following the \gls{3gpp} documentation \cite{3gppchannel}, we model the system as a sector of a three-sector hexagonal cell, where the users are fixed and uniformly distributed within the sector. The cell radii for \gls{rma} and \gls{uma} are obtained from Tables~7.2-1 and 7.2-3 in \cite{3gppchannel} and are around 1000~m and 300~m, respectively. Similar to \cite{Andrew-2022, Joao-2025}, we adopt a \gls{3gpp}-based tapped-delay-line channel model containing 8 taps based on Table 7.7.2-5 from \cite{3gppchannel} where the first tap can be either \gls{los} or \gls{nlos}, and the remaining taps are \gls{nlos}. The channel model has all key practical components, such as path loss and shadowing, small-scale fading, and transmit and receive correlation based on an exponential model (see \cite{Andrew-2022} and references therein for the exact channel formulation). The channel parameters are taken from Tables 7.2-1, 7.2-3, 7.4.1-1, 7.4.2-1, 7.5-6 in \cite{3gppchannel}. 

The 5G \gls{mcs} mapping employed in this work is shown in Table~\ref{tab:MCS} and is taken from \cite{3gppmcs}. As discussed in Section~\ref{sec:gus}, we fit the \gls{mcs}-based rate function with a function of the form $A\log(1 + D\gamma)$, with coefficients $D = 0.5191$ and $A = 1.389$.

\begin{table*}[t] 
    \centering
    \caption{5G MCS mapping \cite{3gppmcs}.}
    \begin{tabular}{|l|ccccccccccccccc|}
        \hline
        MCS $l$ & 1 & 2 & 3 & 4 & 5 & 6 & 7 & 8 & 9 & 10 & 11 & 12 & 13 & 14 & 15 \\
        \hline
        $\zeta_l$ (rate in bps/Hz) & 0.15 & 0.38 & 0.88 & 1.48 & 1.91 & 2.41 & 2.73 & 3.32 & 3.90 & 4.52 & 5.12 & 5.55 & 6.23 & 6.91 & 7.40\\
        \hline
        $\Gamma_l$ (SNR in dB) & -6.82 & -3.44 & -0.53 & 3.79 & 5.80 & 8.08 & 9.76 & 11.72 & 13.49 & 15.87 & 17.73 & 19.50 & 21.32 & 23.51 & 25.15 \\
        \hline
    \end{tabular}
    \label{tab:MCS}
\end{table*}

To obtain robust results and conclusions, we have to consider the randomness in the channel and user positions within the cell. For that, we introduce the concept of realization. We define a \emph{realization} $\Omega (\Delta)$ over a horizon of $\Delta$ \glspl{ts} as a set of fixed users $\mathcal{U}$ with their channel matrices for all $\Delta C$ \glspl{prb}. Within a realization, the varying channel element for a user is the small-scale fading, which changes independently across distinct reporting blocks consisting of $C_B$ subchannels and $T_B$ \glspl{ts}. A reporting block represents a set of \glspl{prb} for which the \gls{csi} of each user is given and constant. We will evaluate many realizations and for each realization, many \glspl{ts} will be executed to allow for the proper evolution of the channels and the weights for \gls{pf}.

Regarding the fairness weights used to compute WSR for a \gls{ts}, in line with the literature \cite{Haseen-2024}, let $R^t_u$ be the average per-\gls{ts} rate seen by user $u \in \mathcal{U}$ in the past $W$ \glspl{ts} at the beginning of \gls{ts} $t$ where $W$ is a parameter to decide on which horizon fairness needs to be enforced. The weight of user $u$ at the beginning of \gls{ts} $t$ is simply $w_u^t=1/R^t_u$. At the end of TS $t$, we have $$R^{t+1}_u = \frac{(W-1)R^{t}_u + r^{t}_u}{W}, \;\; \forall u \in \mathcal{U},$$ where $r^{t}_u$ is the rate transmitted by user $u$ in TS $t$ and the weights are updated accordingly \cite{Haseen-2024, Chen-2021}.

For each realization $\Omega(\Delta)$, given that we use PF, the metric of interest is the \gls{gm} of the rates seen by the users in the $\Delta$ TSs \cite{Haseen-2024}, i.e., 
$$GM(\Delta)= (\prod_{u \in \mathcal{U}} \sum_{t=1}^\Delta r^{t}_u)^{1/|\mathcal{U}|}$$
Hence, the computations for each realization run for $\Delta$ \glspl{ts}. 


\begin{table}[t]
\begin{center}
\caption{Parameters (RB: reporting block, $\#$: number).}
\footnotesize
\begin{tabular}{|l|c|}
\hline
 \textbf{Parameter} & \textbf{Value} \\ 
\hline
Maximum $\#$ of TSs per realization $\Delta$ & 66 \\
\hline
Window length in TSs ($W$) & 6 \\
\hline
$\#$ of channels ($C_B$) and TSs ($T_B$) in a RB & 13 and 2\\
\hline
Subchannels per TS ($C$)  & 78 \\
\hline
Carrier frequency ($f_c$) & 3.5 GHz \\
\hline
Subchannel bandwidth ($B_C$) & 360 kHz \\
\hline
Antennas at the BS ($M_{B}$);  at UE ($M_U$) & 100, 64; 4, 2, 1\\
\hline
Power budget per user in mW ($P_U$) & 50 (RMa); 1 and 5 (UMa)\\
\hline
GUS stopping parameter ($\beta$) & $1.05$ \\
\hline
Channel correlation (BS and users) & 0.4 \\
\hline
Adjacent antenna distance (BS and users) & $0.5 \times 3 \times 10^8 / f_c$ \\
\hline
Noise power density & -174 dBm/Hz \\ 
\hline
Noise figure & 9 dB \\ 
\hline
Coefficients for MCS fitting $A$; $D$ & 1.389; 0.5191 \\
\hline
\end{tabular}
\label{tab:param}
\end{center}
\end{table}

The values of the other parameters for the evaluations are shown in Table~\ref{tab:param}. The system bandwidth is set to 30 MHz, which yields 78 subchannels per \gls{ts} \cite{3gpp-bandwidth}, each with a bandwidth of 360~kHz following the \gls{nr} numerology~1 \cite{3gpp-nr}. We set the \gls{gus} stopping parameter $\beta$ to 1.05 (higher values for $\beta$ were tested without performance improvement) and set the window $W=6$.

The evaluations are conducted using \gls{gus} equipped with the rate-reusing method, except when indicated otherwise.   The results are averaged over 50 realizations for a given number of users, and confidence intervals are computed using the $t$-distribution with a confidence level of 90\%.

We will start our evaluation with a comparison of the performance of the \gls{zf} \gls{bf} strategies \gls{ctrf}, \gls{bd} and \gls{ctrone} in both \gls{3gpp} scenarios for different system parameters. We then study the impact of power management and end this section by presenting the run time reduction in our \gls{gus} heuristic when using the proposed rate-reusing method.

\subsection{ZF BF Strategies Evaluation}

Fig.~\ref{fig:RMa} (top figure) presents, for the \gls{rma} scenario, the GM rates per TS for \gls{ctrf}, \gls{bd} and \gls{ctrone} as a function of the number of users for $M_B=100$ and $M_B=64$ antennas at the \gls{bs}. The number of antennas, $M_U$, at the users is 4 and $P_U = 50$~mW. The ratios in the \gls{gm} rates between \gls{bd} and \gls{ctrf}, and between \gls{ctrone} and \gls{ctrf} are also shown in Fig.~\ref{fig:RMa} (bottom figure). The main observation is that \gls{bd} is not always better than \gls{ctrone}. Their relative performance depends on the values of $M_B$ and $|\mathcal{U}|$. Typically, \gls{bd} is significantly better for a small number of users while the reverse is true for a large number of users. Clearly, as expected, \gls{ctrf} outperforms the other strategies across all numbers of users and \gls{bs} antennas and the performance gap can be significant. More \gls{bs} antennas reduce the gap between \gls{bd} and \gls{ctrf}, while it is the reverse for \gls{ctrone}. We can also see in the top figure that $M_B$ has a significant impact on performance for the three strategies for all values of the number of users. Specifically, for \gls{ctrf} and \gls{bd}, the improvement in the \gls{gm} rates for $M_B=100$ relative to $M_B=64$ starts at around 30\% for 10 users and grows with the number of users to above 40\%. For \gls{ctrone}, similar trends are observed, except that the improvement for 10 users is smaller at roughly 16\%.

These observations can be understood better by referring to Fig. \ref{fig:CombStreamRMa}, which shows the empirical distribution of the stream allocation pattern per selected user per \gls{prb} given by \gls{ctrf} for different numbers of users and $M_B=64$ \gls{bs} antennas. In Fig. \ref{fig:CombStreamRMa}, the $x$ axis represents which stream pattern is enabled per scheduled user per \gls{prb}. Recall that the streams are numbered from 1 to $M_U=4$ in order of strength. For example, referring to the $|\mathcal{U}|=10$ case, 30\% of the users have only stream 1 scheduled in a PRB allocated to them, and around 20\% have streams 1, 2, and 3. For $|\mathcal{U}|=10$ users, around 40\% of the time, at least three streams are enabled per scheduled user, whereas for $|\mathcal{U}|=40$ users, users with high stream counts barely occur and one-stream users are observed in almost 80\% of the cases. These results are in line with the performance given by Fig. \ref{fig:RMa}, where \gls{ctrone}, which only uses one stream per scheduled user, performs comparably to \gls{ctrf} for 40 users, and \gls{bd}, presenting four streams per scheduled user, does similarly for 10 users.

\begin{figure}[t]
\centering
\includegraphics[width=0.45\textwidth]{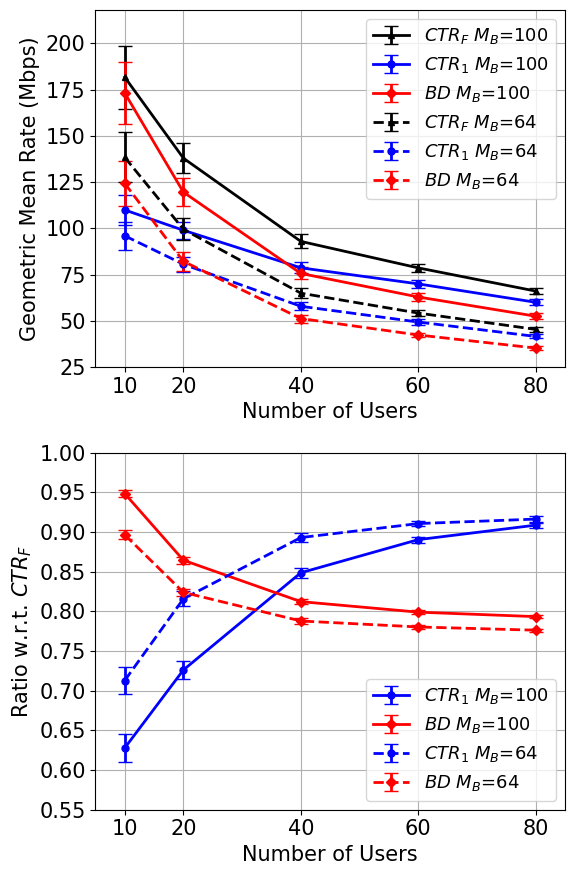}
\caption{\gls{rma} scenario: $M_B=100$ and $M_B=64$ antennas at the BS, $M_U=4$ antennas at the users, $P_U = 50$~mW. Top figure: \gls{gm} rates in Mbps per \gls{ts} vs. number of users for \gls{ctrf}, \gls{bd} and \gls{ctrone}. Bottom figure: ratios in the \gls{gm} rates between \gls{bd} and \gls{ctrf}, and between \gls{ctrone} and \gls{ctrf}.}
\label{fig:RMa}
\end{figure}

\begin{figure}[t]
\centering
\includegraphics[width=0.47\textwidth]{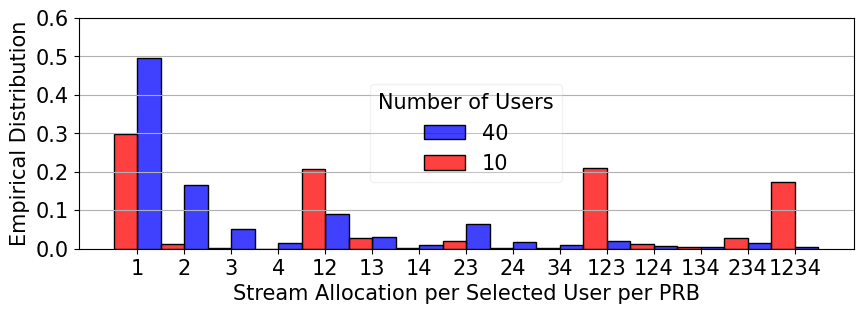}
\caption{\gls{rma} scenario: empirical distribution of the stream allocation per scheduled user per PRB given by \gls{ctrf} for $|\mathcal{U}| \in \{10,40\}$ users with $M_B=64$ \gls{bs} antennas,  $M_U=4$ antennas at the users, and $P_U = 50$~mW.}
\label{fig:CombStreamRMa}
\end{figure}

Next, we show the equivalent results for \gls{uma} in Figs.~\ref{fig:UMa} and \ref{fig:CombStreamUMa} when $P_U = 1$~mW. Compared to the results for \gls{rma} (see Fig. \ref{fig:RMa}), the main differences are that the gap between \gls{ctrone} and \gls{ctrf} is small across all values of the number of users for the two values of $M_B$ and that \gls{ctrone} outperforms \gls{bd}, except for $|\mathcal{U}|=10$ and $M_B=100$. These observations are corroborated by Fig.~4, where the most common stream allocation given by \gls{ctrf} is stream~1 enabled per scheduled user, which is essentially \gls{ctrone}. Similar to the \gls{rma} case, we can also observe in the top figure that $M_B$ also has a considerable impact on performance for the three strategies for all values of the number of users. Particularly, for all strategies, the improvement in the \gls{gm} rates with $M_B=100$ compared to $M_B=64$ is at least 30\% for a few users (e.g., 10 users) and increases with the number of users to reach over 40\%.

\begin{figure}[t]
\centering
\includegraphics[width=0.45\textwidth]{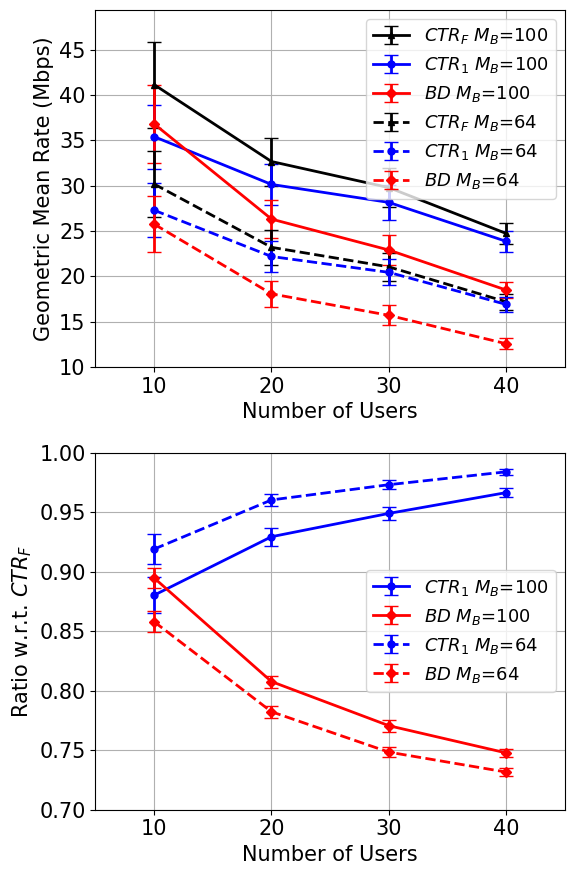}
\caption{\gls{uma} scenario: $M_B=100$ and $M_B=64$ antennas at the \gls{bs}, $M_U=4$ antennas at the users, and $P_U = 1$~mW. Top figure: \gls{gm} rates in Mbps per \gls{ts} vs. number of users for \gls{ctrf}, \gls{bd} and \gls{ctrone}. Bottom figure: ratios in the \gls{gm} rates between \gls{bd} and \gls{ctrf}, and between \gls{ctrone} and \gls{ctrf}.}
\label{fig:UMa}
\end{figure}

\begin{figure}[t]
\centering
\includegraphics[width=0.47\textwidth]{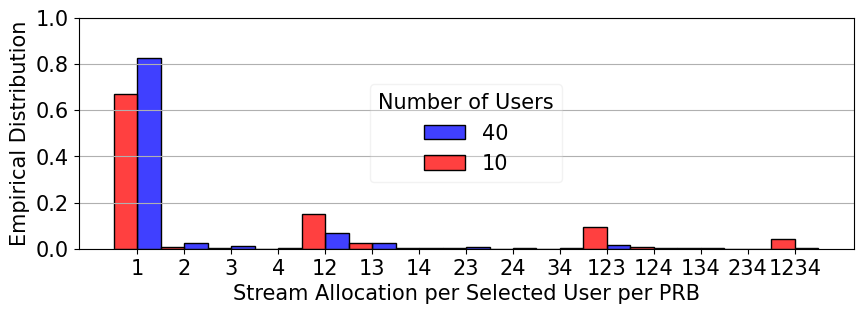}
\caption{\gls{uma} scenario: empirical distribution of the stream allocation per selected user per PRB given by \gls{ctrf} for $|\mathcal{U}| \in \{10,40\}$ users with $M_B=64$ \gls{bs} antennas,  $M_U=4$ antennas at the users, and $P_U = 1$~mW.}
\label{fig:CombStreamUMa}
\end{figure}

In summary, depending on the \gls{3gpp} scenario, the strategy that could replace \gls{ctrf},  if deemed too complex to operate in real-time, varies. For \gls{rma}, considering that the number of users is generally not large (below 30) in a sector, \gls{bd} could replace the more complex \gls{ctrf} given its comparable performance. For larger user counts, \gls{ctrone} could be an alternative to \gls{ctrf} because of the small performance gap. For \gls{uma}, \gls{ctrone} emerges as a suitable option over \gls{ctrf} due to its lower complexity and similar performance.

Next, we evaluate the performance of the three \gls{zf} \gls{bf} strategies for different numbers of antennas, $M_U$, at the users in the \gls{uma} scenario. For this study, we employ a per-user power budget of $P_U = 5$~mW because for $M_U=1$, $P_U = 1$~mW can yield unconnected users. Fig.~\ref{fig:UMaMU} (top figure) presents the \gls{gm} rates per \gls{ts} for the \gls{zf} strategies vs. number of users for different values of $M_U$ with $M_B=64$ \gls{bs} antennas. Fig.~\ref{fig:UMaMU} (bottom figure) shows the corresponding ratios for \gls{bd} and \gls{ctrone} relative to \gls{ctrf}. For $M_U=1$, only results for \gls{ctrone} are provided, given that all strategies are equivalent. We can observe that more antennas at the users produce a significant improvement in the \gls{gm} rates delivered by \gls{ctrf} and \gls{ctrone}. Specifically, the improvements in the \gls{gm} rates for \gls{ctrf} (from $M_U=2$ to $M_U=4$) and \gls{ctrone} (from $M_U=1$ to $M_U=2$ and from $M_U=2$ to $M_U=4$) fluctuate at approximately 20\% for all numbers of users. Conversely, the improvement for \gls{bd} is low for a small number of users and is nonexistent when the number of users increases. For instance, for 10 users, the improvement from $M_U=2$ to $M_U=4$ for \gls{bd} is around 12\% but drops to nearly zero for 40 users. Another interesting observation is that the performance disparities for \gls{ctrone} and \gls{bd} w.r.t. \gls{ctrf} decrease for $M_U=2$ compared to $M_U=4$. This can be explained by the fact that with fewer antennas at the users, \gls{ctrf} has less flexibility in the stream allocation and loses part of its potential.

\begin{figure}[t]
\centering
\includegraphics[width=0.45\textwidth]{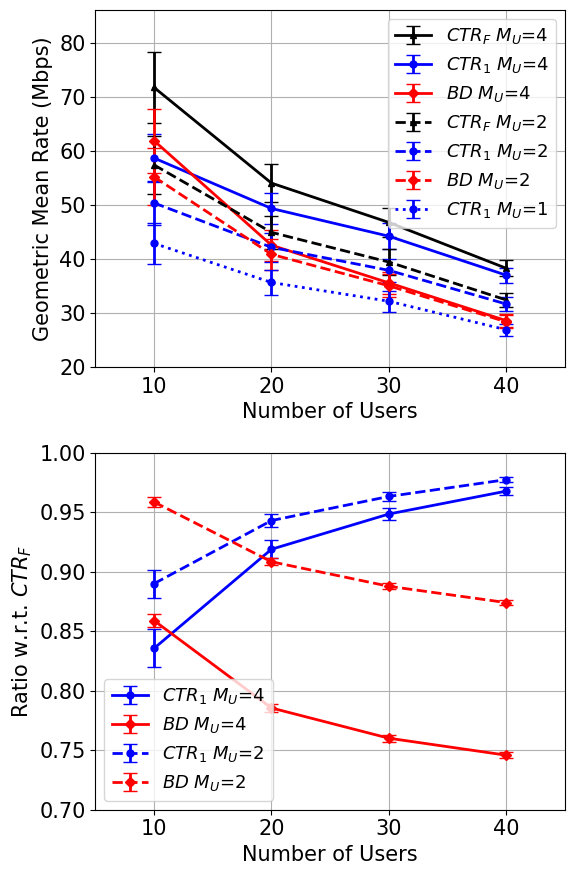}
\caption{\gls{uma} scenario: $M_B=64$ antennas at the \gls{bs}, $M_U=4$, $M_U=2$ and $M_U=1$ antennas at the users, $P_U=5$~mW. Top figure: \gls{gm} rates in Mbps per \gls{ts} vs. number of users for \gls{ctrf}, \gls{bd} and \gls{ctrone}. Bottom figure: ratios in the \gls{gm} rates between \gls{bd} and \gls{ctrf}, and between \gls{ctrone} and \gls{ctrf}.}
\label{fig:UMaMU}
\end{figure}

We also evaluate the performance of the \gls{zf} \gls{bf} strategies for different per-user power budgets, $P_U$. Fig.~\ref{fig:UMaP} (top figure) presents the \gls{gm} rates per \gls{ts} for the \gls{zf} strategies vs. number of users for different values of $P_U$ with $M_B=64$ \gls{bs} antennas, $M_U = 4$ antennas at the users in the \gls{uma} scenario. Fig.~\ref{fig:UMaP} (bottom figure) shows the corresponding ratios for \gls{bd} and \gls{ctrone} relative to \gls{ctrf}. The \gls{gm} rates for all \gls{zf} \gls{bf} strategies approximately double  for $P_U=5$~mW compared to $P_U=1$~mW, which is very significant. Also, the performance gap for \gls{bd} w.r.t. \gls{ctrf} remains nearly the same for the two power budgets. On the other hand, the difference between \gls{ctrone} and \gls{ctrf} increases with more power. One reason for this is that, owing to the extra capacity provided by increased power, more streams per scheduled user can be enabled with \gls{ctrf}.

\begin{figure}[t]
\centering
\includegraphics[width=0.45\textwidth]{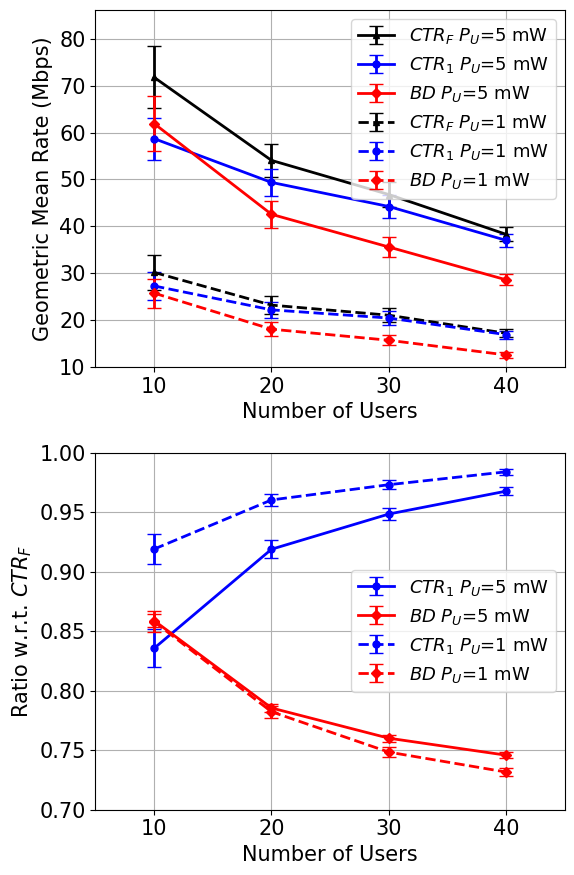}
\caption{\gls{uma} scenario: $M_B=64$ antennas at the \gls{bs}, $M_U=4$ antennas at the users, $P_U=5$~mW and $P_U=1$~mW. Top figure: \gls{gm} rates in Mbps per \gls{ts} vs. number of users for \gls{ctrf}, \gls{bd} and \gls{ctrone}. Bottom figure: ratios in the \gls{gm} rates between \gls{bd} and \gls{ctrf}, and between \gls{ctrone} and \gls{ctrf}. }
\label{fig:UMaP}
\end{figure}

\subsection{Power Management Schemes Evaluation}
All the results presented above were obtained with the complex power management defined in Section~\ref{sec:gus} as \gls{tpm}. Next, we study how a simpler power management scheme, called \gls{epm}, performs for each \gls{zf} \gls{bf} strategy. This scheme allocates, for each user,  the power budget equally among its selected streams over the \gls{ts}. This scheme is used during and at the end of the greedy-up search explained in Section~\ref{sec:gus}.

\begin{figure}[t]
\centering
\includegraphics[width=0.45\textwidth]{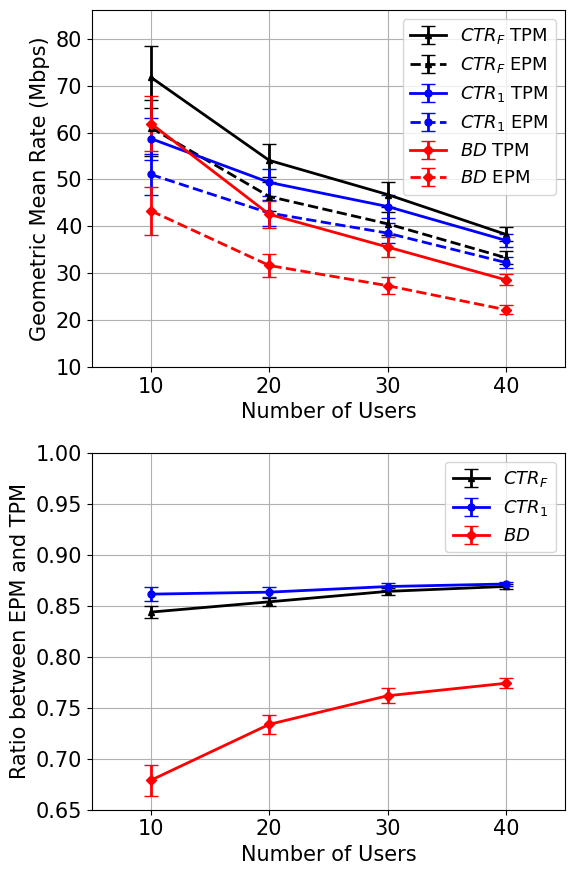}
\caption{\gls{uma} scenario: $M_B=64$ antennas at the \gls{bs}, $M_U=4$ antennas at the users, $P_U=5$~mW. Top figure: \gls{gm} rates in Mbps per \gls{ts} vs. number of users for \gls{ctrf}, \gls{bd} and \gls{ctrone} for the power management schemes \gls{tpm} and \gls{epm}. Bottom figure: ratio in the \gls{gm} rates between \gls{epm} and \gls{tpm} for each \gls{zf} \gls{bf} strategy.}
\label{fig:TPMvsEPM}
\end{figure}

Fig.~\ref{fig:TPMvsEPM} (top figure) presents, for the \gls{uma} scenario, the \gls{gm} rates per \gls{ts} for the three \gls{zf} strategies vs. number of users for \gls{tpm} and \gls{epm} with $M_B=64$ \gls{bs} antennas, $M_U=4$ antennas at the users, and a per-user power budget of $P_U = 5$~mW. Fig.~\ref{fig:TPMvsEPM} (bottom figure) shows the corresponding ratio between \gls{epm} and \gls{tpm} for each \gls{zf} \gls{bf} strategy. The performance drop for \gls{bd} using the simpler \gls{epm} is larger compared to the one for \gls{ctrone} or \gls{ctrf}. For example, the ratio between \gls{epm} and \gls{tpm} for \gls{bd} is just above 75\% for 40 users but can reach nearly as low as 65\%. In contrast, the ratios for \gls{ctrone} and \gls{ctrf} are approximately 85\% for all numbers of users. An explanation for that is that, with \gls{bd}, all $M_U$ streams are enabled per scheduled users per \gls{prb}, and these streams have distinct strengths. \gls{tpm} distributes the power (i.e., \gls{pd}) taking into account the streams' gains, while that does not occur with \gls{epm}. \gls{ctrone} and \gls{ctrf} do not suffer from this issue because in the former, only the strongest stream is enabled, while the latter allows a flexible stream allocation. In conclusion, in case a simpler power management scheme, such as \gls{epm}, is required for real-time settings, the outcomes suggest that \gls{ctrf} or \gls{ctrone} are better suited but the performance cost is not negligible.

\subsection{Miscellaneous}
Finally, we investigate how much decrease in run time we can obtain by executing \gls{gus} with the more aggressive rate-reusing method proposed in Section~\ref{sec:reusing} w.r.t. \gls{gus} without it as proposed in Section~\ref{sec:search}. We consider   the \gls{rma} scenario with $M_B=64$ \gls{bs} antennas, $|\mathcal{U}| \in \{20,40,60,80\}$ users, $M_U=4$ antennas at the users, and $P_U=50$~mW. We use \gls{ctrone} and \gls{bd} to illustrate that. We do not employ \gls{ctrf} as an example, since it can take dozens of hours for just one realization, even with the rate-reusing method. The complexity of \gls{ctrf} comes from the increased granularity in the stream selection process (i.e., flexible stream allocation per user) and the fact that the \gls{zf} \gls{bf} computations are more costly. The run times are averaged over at least 650 \glspl{ts} for a given number of users. For \gls{ctrone}, the decrease in run time is up to roughly 11\% and for \gls{bd}, the run time reduction ranges approximately from 26\% to 42\%. Computational efficiency is critical due to the long period needed to conduct the comprehensive evaluations in this work. For instance, more than four hours were required for only one realization of \gls{ctrone} with 80 users without the rate-reusing method. In this study, we had to execute a large number of realizations for different \gls{zf} \gls{bf} strategies, across different \gls{3gpp}-based scenarios and system parameters, highlighting the importance of the proposed rate-reusing method for run time reduction.

\section{Conclusion} \label{sec:conclusion}
This study is the first of its kind. Very few papers have addressed the uplink of an \gls{mu-mimo}/\gls{ofdma} single cell  with multi-antenna users and multiple channels. Of those, none have considered  different \gls{zf} \gls{bf} strategies and all \gls{rrm} processes. We have evaluated  the performance that can be achieved on the uplink of an \gls{mu-mimo}/\gls{ofdma} single cell  with multi-antenna users for three \gls{zf} \gls{bf} strategies: \gls{bd}, where all possible data streams are enabled per scheduled user; \gls{ctrone}, in which only the strongest streams of the scheduled users are used; \gls{ctrf}, which allows a flexible stream allocation per user. \gls{ctrf} offers potentially better performance due to its flexibility, albeit at the price of higher complexity compared to other strategies. This study was based on a system with proportional fairness,  practical \glspl{mcs} for \gls{3gpp}-based scenarios (\gls{rma} and \gls{uma}). The \gls{ul} \gls{rrm} must be done over the entire \gls{ts} at once due to the \gls{ul} power management and many \glspl{ts} must be executed to take fairness into account. Hence, there is a need for efficient tools to evaluate the performance of such a system. We have proposed a heuristic based on a greedy-up search for stream-sets that operates over the \gls{ts} and considers an \glspl{mcs}-based rate function, all key \gls{rrm} processes and \gls{pf}. To make the heuristic tractable for this study, we equipped it with efficient \gls{zf} \gls{bf} and power computation methods and proposed a novel rate-reusing method where rates computed in previous search iterations are reused whenever possible.

The results and discussions presented in Section~\ref{sec:results} provide valuable information that can guide the design of real-time \gls{ul} \gls{rrm} heuristics, which is an important avenue for future research. Specifically, the results showed that \gls{ctrf} outperformed \gls{bd} and \gls{ctrone}, but in some cases, \gls{bd} or \gls{ctrone} could replace \gls{ctrf} in real-time settings given their comparable performance and lower complexity. For instance, for \gls{rma}, either \gls{bd} or \gls{ctrone} could be an alternative to \gls{ctrf} depending on the number of users, while for \gls{uma}, \gls{ctrone} performed similarly to \gls{ctrf} across all system parameters. We also showed that the numbers of antennas at the \gls{bs} and at the users, and the power budget per user considerably impacted the performance of the \gls{zf} strategies. Finally, we compared the performance of each \gls{zf} \gls{bf} strategy with a complex power management scheme to that with a simpler scheme (based on equal power management), where we saw that the \gls{bd} performance was the most impacted by the simpler scheme relative to the other \gls{zf} \gls{bf} strategies. Another avenue for future work is to consider a multi-cell setting, where the interference from neighbouring cells must be taken into account.

\bibliographystyle{IEEEtran}
\bibliography{main}

\appendices 
\section{Rate-reusing Rules for BD}
\label{append}

At the end of an iteration, we have a newly selected candidate, and we will classify the candidates for the next iteration into four types (some types are exclusive and some are not) w.r.t. the newly selected candidate. We are at the beginning of an iteration, and the selected candidate at the previous iteration was in PRB $\hat{c}$, which we call the reference \gls{prb}. Recall that for a candidate $(u,c)$, according to Section \ref{sec:search}, we perform rate computations through \gls{pa} and \gls{pd} for users that have already been selected in \gls{prb} $c$ as well as for user $u$, and we can reuse the rates of the other users. What we are trying to do next is limit further the number of those users for whom we do indeed perform \gls{pa} and \gls{pd} for rate computations and instead use rate computations done earlier. The candidate types are:

\begin{itemize}
\item \textbf{Type 1}: This type includes all the candidates $(u,\hat{c})$.   These candidates  cannot be classified into other types and no further rate reuse can be done other than the ones defined in the rate computations of Section \ref{sec:search}. 
    
\item \textbf{Type 2}: This type includes all candidates with a user already selected in the reference \gls{prb}, i.e., all candidates $(u,c)$ such that user $u$ has been selected in \gls{prb} $\hat{c}$ (note that $c \neq \hat{c}$ by definition). To compute the \gls{wsr} over the \gls{ts} for each such candidate, we have to execute one rate computation through \gls{pa} and \gls{pd} for the candidate user $u$ to obtain its rates. The rates of the other users selected in the other \glspl{prb}, excluding the reference \gls{prb}, can be reused from previous iterations. The rates of the users selected in the reference \gls{prb} (other than the candidate user $u$) might have to be updated. This last case occurs when the candidate $(u,c)$ is not only a candidate of type 2 but also type 3. Type 3 candidates are discussed below. 

To compute the \gls{wsr} $WSR^c_u$ for a candidate $(u,c)$ of type 2, assuming we temporarily select user $u$ in \gls{prb} $c$, we execute a rate computation by doing a \gls{pa} and \gls{pd} (through $\overline{\text{JPM}}$) for user $u$ and obtain the rates $r^b_{u,j}, \forall b \in \mathcal{C}, \forall j \in \mathcal{M}_U$ for its selected streams. Whether we can reuse or have to compute the rates for users $k \in \mathcal{U}$ selected in \gls{prb} $\hat{c}$ with $k \neq u$ is tackled below for candidates of type 3. 
    
\item \textbf{Type 3}: Candidates of type 3 are characterized as follows: consider the set $\mathcal{K}$ of all users that have been selected in the reference \gls{prb} $\hat{c}$. Let $\hat{\mathcal{C}}_k \subseteq \mathcal{C}$ be the set of \glspl{prb} for which user $k \in \mathcal{K}$ has been selected. Candidates $(u,c)$, with $c \neq \hat{c}$, such that $ \exists k \in \mathcal{K}$ with $c \in \hat{\mathcal{C}}_k$ is a candidate of type 3, for which we have to execute one rate computation through \gls{pa} and \gls{pd} for each $k \in \mathcal{K}$ such that $c \in \hat{\mathcal{C}}_k$. If the candidate is also of type 2, an extra rate computation for user $u$ is required over its selected \glspl{prb} plus \gls{prb} $c$ as seen previously.
To compute the \gls{wsr} $WSR^c_u$ for a candidate $(u,c)$ of type 3, for each user $k \in \mathcal{K}$ such that $c \in \hat{\mathcal{C}}_k$, we execute a rate computation by \gls{pa} and \gls{pd} (through $\overline{\text{JPM}}$) and obtain the rates $r^b_{k,j}, \forall b \in \mathcal{C}, \forall s \in \mathcal{M}_U$ for its selected streams.

\item \textbf{Type 4}: Candidates of type 4 do not require new rate computations for the \gls{wsr} computation. A type-4 candidate is any candidate that is not categorized into any of the previously described types. In this candidate type, we can reuse all the rates computed in previous search iterations.
\end{itemize}

A candidate can be of types 2 and 3 simultaneously, as discussed, but candidates of type 1 or 4 cannot be of other types. The reuse of rates for all candidate types in a given search iteration relies on the assumption that we computed those rates previously in some search iteration. We have presented above the rules governing which rates have to be computed and which rates can be reused for \gls{wsr} computation in a given search iteration depending on the candidate type for \gls{bd}.

\end{document}

%% file: glossary.tex
\newacronym{mu-mimo}{MU-MIMO}{Multi-User Multiple-Input Multiple-Output}
\newacronym{su-mimo}{SU-MIMO}{Single-User Multiple-Input Multiple-Output}
\newacronym{ofdma}{OFDMA}{Orthogonal Frequency-Division Multiple Access}
\newacronym{bd}{BD}{Block Diagonalization}
\newacronym{ctrf}{CTR\textsubscript{F}}{Coordinated-Transmit-Receive-Flexible}
\newacronym{ctrone}{CTR\textsubscript{1}}{Coordinated-Transmit-Receive-1}
\newacronym{mcs}{MCS}{Modulation and Coding Scheme}
\newacronym{3gpp}{3GPP}{3rd Generation Partnership Project}
\newacronym{pf}{PF}{Proportional Fairness}
\newacronym{dl}{DL}{Downlink}
\newacronym{ul}{UL}{Uplink}
\newacronym{bs}{BS}{Base-Station}
\newacronym{met}{MET}{Multiple Eigenmode Transmission}
\newacronym{mdet}{MDET}{Multiple Dominant Eigenmode Transmission}
\newacronym{gus-bd}{GUS-BD}{Greedy User Selection - Block Diagonalization}
\newacronym{prb}{PRB}{Physical Resource Block}
\newacronym{zf}{ZF}{Zero-Forcing}
\newacronym{snr}{SNR}{Signal-to-Noise Ratio}
\newacronym{wsr}{WSR}{Weighted Sum-Rate}
\newacronym{rma}{RMa}{Rural Macro}
\newacronym{uma}{UMa}{Urban Macro}
\newacronym{umi}{UMi}{Urban Micro}
\newacronym{fwa}{FWA}{Fixed Wireless Access}
\newacronym{csi}{CSI}{Channel State Information}
\newacronym{ts}{TS}{Time-Slot}
\newacronym{tdd}{TDD}{Time-Division Duplexing}
\newacronym{los}{LOS}{Line-Of-Sight}
\newacronym{nlos}{NLOS}{Non-Line-Of-Sight}
\newacronym{svd}{SVD}{Singular Value Decomposition}
\newacronym{ctr}{CTR}{Coordinated Transmit-Receive}
\newacronym{fwa-rma}{FWA-RMa}{Fixed Wireless Access - Rural Macro}
\newacronym{r-rma}{R-RMa}{Regular - Rural Macro}
\newacronym{wf}{WF}{Water-Filling}
\newacronym{ma}{MA}{Multi-Antenna}
\newacronym{sr}{SR}{Sum-Rate}
\newacronym{rrm}{RRM}{Radio Resource Management}
\newacronym{pa}{PA}{Power Allocation}
\newacronym{pd}{PD}{Power Distribution}
\newacronym{dpc}{DPC}{Dirty Paper Coding}
\newacronym{ml}{ML}{Machine Learning}
\newacronym{gnn}{GNN}{Graph Neural Network}
\newacronym{sic}{SIC}{Successive Interference Cancellation}
\newacronym{mmse}{MMSE}{Minimum Mean Square Error}
\newacronym{sa}{SA}{Single-Antenna}
\newacronym{mrt}{MRT}{Maximum Ratio Transmission}
\newacronym{mrc}{MRC}{Maximum Ratio Combining}
\newacronym{rf}{RF}{Radio Frequency}
\newacronym{rq}{RQ}{Research Question}
\newacronym{tm}{TM}{Target Metric}
\newacronym{us}{US}{User Selection}
\newacronym{bf}{BF}{Beamforming}
\newacronym{mu}{MU}{Multi-User}
\newacronym{mr}{MR}{Maximum Ratio}
\newacronym{gus}{GUS}{Greedy-Up Search}
\newacronym{gds}{GDS}{Greedy-Down Search}
\newacronym{epm}{EPM}{Equal Power Management}
\newacronym{tpm}{TPM}{Two-Step Power Management}
\newacronym{ss}{SS}{Stream Selection}
\newacronym{bler}{BLER}{Block-Error Rate}
\newacronym{gm}{GM}{Geometric Mean}
\newacronym{nr}{NR}{New Radio}
\newacronym{qos}{QoS}{Quality-Of-Service}

%% file: main.bbl
\begin{thebibliography}{10}
\providecommand{\url}[1]{#1}
\csname url@samestyle\endcsname
\providecommand{\newblock}{\relax}
\providecommand{\bibinfo}[2]{#2}
\providecommand{\BIBentrySTDinterwordspacing}{\spaceskip=0pt\relax}
\providecommand{\BIBentryALTinterwordstretchfactor}{4}
\providecommand{\BIBentryALTinterwordspacing}{\spaceskip=\fontdimen2\font plus
\BIBentryALTinterwordstretchfactor\fontdimen3\font minus \fontdimen4\font\relax}
\providecommand{\BIBforeignlanguage}[2]{{%
\expandafter\ifx\csname l@#1\endcsname\relax
\typeout{** WARNING: IEEEtran.bst: No hyphenation pattern has been}%
\typeout{** loaded for the language `#1'. Using the pattern for}%
\typeout{** the default language instead.}%
\else
\language=\csname l@#1\endcsname
\fi
#2}}
\providecommand{\BIBdecl}{\relax}
\BIBdecl

\bibitem{Marzetta-2015}
T.~L. Marzetta, ``Massive \uppercase{MIMO}: An introduction,'' \emph{Bell Labs Technical Journal}, vol.~20, pp. 11--22, 2015.

\bibitem{Castaneda-2017}
E.~Castañeda, A.~Silva, A.~Gameiro, and M.~Kountouris, ``An overview on resource allocation techniques for multi-user \uppercase{MIMO} systems,'' \emph{IEEE Communications Surveys \& Tutorials}, vol.~19, no.~1, pp. 239--284, 2017.

\bibitem{Bjornson-2013}
E.~Björnson, M.~Kountouris, M.~Bengtsson, and B.~Ottersten, ``Receive combining vs. multi-stream multiplexing in downlink systems with multi-antenna users,'' \emph{IEEE Transactions on Signal Processing}, vol.~61, no.~13, pp. 3431--3446, 2013.

\bibitem{Haseen-2024}
H.~Rahman and C.~Rosenberg, ``Resource allocation for the uplink of a multi-user massive \uppercase{MIMO} system,'' \emph{IEEE Transactions on Mobile Computing}, vol.~24, no.~5, pp. 4326--4338, 2025.

\bibitem{Ozcan-2021}
Y.~Özcan and C.~Rosenberg, ``Uplink scheduling in multi-cell \uppercase{OFDMA} networks: A comprehensive study,'' \emph{IEEE Transactions on Mobile Computing}, vol.~20, no.~10, pp. 3081--3098, 2021.

\bibitem{Boccardi-2007}
F.~Boccardi and H.~Huang, ``A near-optimum technique using linear precoding for the \uppercase{MIMO} broadcast channel,'' in \emph{2007 IEEE International Conference on Acoustics, Speech and Signal Processing - ICASSP '07}, vol.~3, 2007, pp. III--17--III--20.

\bibitem{Spencer-2004}
Q.~Spencer, A.~Swindlehurst, and M.~Haardt, ``Zero-forcing methods for downlink spatial multiplexing in multiuser \uppercase{MIMO} channels,'' \emph{IEEE Transactions on Signal Processing}, vol.~52, no.~2, pp. 461--471, 2004.

\bibitem{Wang-2008}
J.~Wang, D.~J. Love, and M.~D. Zoltowski, ``User selection with zero-forcing beamforming achieves the asymptotically optimal sum rate,'' \emph{IEEE Transactions on Signal Processing}, vol.~56, no.~8, pp. 3713--3726, 2008.

\bibitem{Shi-2008}
Z.~Shi, C.~Zhao, and Z.~Ding, ``Low complexity eigenmode selection for \uppercase{MIMO} broadcast systems with block diagonalization,'' in \emph{2008 IEEE International Conference on Communications}, 2008, pp. 3976--3981.

\bibitem{Tran-2012-1}
L.-N. Tran, M.~Bengtsson, and B.~Ottersten, ``Iterative precoder design and user scheduling for block-diagonalized systems,'' \emph{IEEE Transactions on Signal Processing}, vol.~60, no.~7, pp. 3726--3739, 2012.

\bibitem{Tran-2012-2}
L.-N. Tran, M.~Juntti, and E.-K. Hong, ``On the precoder design for block diagonalized \uppercase{MIMO} broadcast channels,'' \emph{IEEE Communications Letters}, vol.~16, no.~8, pp. 1165--1168, 2012.

\bibitem{Chen-2008}
R.~Chen, Z.~Shen, J.~G. Andrews, and R.~W. Heath, ``Multimode transmission for multiuser \uppercase{MIMO} systems with block diagonalization,'' \emph{IEEE Transactions on Signal Processing}, vol.~56, no.~7, pp. 3294--3302, 2008.

\bibitem{Joao-2025}
J.~P. P.~G. Marques, K.~Danilchenko, and C.~Rosenberg, ``Performance evaluation of \uppercase{MU-MIMO} systems with multi-antenna users for different precoding strategies,'' in \emph{2025 IEEE Wireless Communications and Networking Conference (WCNC)}, 2025, pp. 1--6.

\bibitem{Shen-2006}
Z.~Shen, R.~Chen, J.~Andrews, R.~Heath, and B.~Evans, ``Low complexity user selection algorithms for multiuser \uppercase{MIMO} systems with block diagonalization,'' \emph{IEEE Transactions on Signal Processing}, vol.~54, no.~9, pp. 3658--3663, 2006.

\bibitem{Hussein-2024}
A.~Hussein, P.~Mitran, and C.~Rosenberg, ``Operating multi-user massive \uppercase{MIMO} networks: Trade-off between performance and runtime,'' \emph{IEEE Transactions on Network and Service Management}, vol.~21, no.~2, pp. 2170--2186, 2024.

\bibitem{Chen-2021}
Y.~Chen, Y.~Wu, Y.~T. Hou, and W.~Lou, ``m\uppercase{C}ore: Achieving sub-millisecond scheduling for \uppercase{5G MU-MIMO} systems,'' in \emph{IEEE INFOCOM 2021 - IEEE Conference on Computer Communications}, 2021, pp. 1--10.

\bibitem{Guthy-2009}
C.~Guthy, W.~Utschick, R.~Hunger, and M.~Joham, ``Weighted sum rate maximization in the \uppercase{MIMO MAC} with linear transceivers: Algorithmic solutions,'' in \emph{2009 Conference Record of the Forty-Third Asilomar Conference on Signals, Systems and Computers}, 2009, pp. 1474--1478.

\bibitem{Andrew-2022}
A.~Lappalainen, Y.~Zhang, and C.~Rosenberg, ``Planning 5\uppercase{G} networks for rural fixed wireless access,'' \emph{IEEE Transactions on Network and Service Management}, vol.~20, no.~1, pp. 441--455, 2023.

\bibitem{Li-2007}
L.~Ping and P.~Wang, ``Multi-user gain and maximum eigenmode beamforming for \uppercase{MIMO} systems with rate constraints,'' in \emph{2007 IEEE Information Theory Workshop on Information Theory for Wireless Networks}, 2007, pp. 1--5.

\bibitem{Feres-2023}
C.~Feres and Z.~Ding, ``An unsupervised learning paradigm for user scheduling in large scale multi-antenna systems,'' \emph{IEEE Transactions on Wireless Communications}, vol.~22, no.~5, pp. 2932--2945, 2023.

\bibitem{Zhang-2024}
L.~Zhang, A.~Liu, and X.~Chen, ``A \uppercase{WMMSE}-based contiguous resource scheduling algorithm for \uppercase{5G-NR} uplink,'' \emph{IEEE Wireless Communications Letters}, vol.~13, no.~2, pp. 466--470, 2024.

\bibitem{Meng-2018}
X.~Meng, Y.~Chang, Y.~Wang, and J.~Wu, ``Multi-user grouping based scheduling algorithm in massive \uppercase{MIMO} uplink networks,'' in \emph{2018 IEEE 4th International Conference on Computer and Communications (ICCC)}, 2018, pp. 409--413.

\bibitem{Zhang-2005}
Y.~J. Zhang and K.~Letaief, ``An efficient resource-allocation scheme for spatial multiuser access in \uppercase{MIMO/OFDM} systems,'' \emph{IEEE Transactions on Communications}, vol.~53, no.~1, pp. 107--116, 2005.

\bibitem{Yi-2011}
X.~Yi and E.~K.~S. Au, ``User scheduling for heterogeneous multiuser \uppercase{MIMO} systems: A subspace viewpoint,'' \emph{IEEE Transactions on Vehicular Technology}, vol.~60, no.~8, pp. 4004--4013, 2011.

\bibitem{Feminias-2016}
G.~Femenias and F.~Riera-Palou, ``Scheduling and resource allocation in downlink multiuser \uppercase{MIMO-OFDMA} systems,'' \emph{IEEE Transactions on Communications}, vol.~64, no.~5, pp. 2019--2034, 2016.

\bibitem{Yuan-2024}
Y.~Quan, S.~Shahsavari, and C.~Rosenberg, ``Planning and operation of millimeter-wave downlink systems with hybrid beamforming,'' \emph{IEEE Transactions on Wireless Communications}, vol.~23, no.~12, pp. 18\,214--18\,227, 2024.

\bibitem{Xing-2020}
C.~Xing, Y.~Jing, S.~Wang, S.~Ma, and H.~V. Poor, ``New viewpoint and algorithms for water-filling solutions in wireless communications,'' \emph{IEEE Transactions on Signal Processing}, vol.~68, pp. 1618--1634, 2020.

\bibitem{3gppchannel}
\BIBentryALTinterwordspacing
3GPP, ``5\uppercase{G}; \uppercase{S}tudy on channel model for frequencies from 0.5 to 100 ghz,'' 3rd Generation Partnership Project, TR 38.901, Version 16.1.0, 2020. [Online]. Available: \url{https://portal.3gpp.org/desktopmodules/Specifications/SpecificationDetails.aspx?specificationId=3173}
\BIBentrySTDinterwordspacing

\bibitem{3gppmcs}
\BIBentryALTinterwordspacing
------, ``5\uppercase{G}; \uppercase{NR}; \uppercase{P}hysical layer procedures for data,'' 3rd Generation Partnership Project, TS 38.214, Version 15.2.0, 2018. [Online]. Available: \url{https://portal.3gpp.org/desktopmodules/Specifications/SpecificationDetails.aspx?specificationId=3216}
\BIBentrySTDinterwordspacing

\bibitem{3gpp-bandwidth}
\BIBentryALTinterwordspacing
------, ``5\uppercase{G}; \uppercase{NR}; \uppercase{U}ser equipment (\uppercase{UE}) radio transmission and reception; \uppercase{P}art 1: Range 1 \uppercase{S}tandalone,'' 3rd Generation Partnership Project, TS 38.101-1, Version 16.2.0, 2020. [Online]. Available: \url{https://portal.3gpp.org/desktopmodules/Specifications/SpecificationDetails.aspx?specificationId=3283}
\BIBentrySTDinterwordspacing

\bibitem{3gpp-nr}
\BIBentryALTinterwordspacing
------, ``5\uppercase{G}; \uppercase{NR}; \uppercase{P}hysical channels and modulation,'' 3rd Generation Partnership Project, TS 38.211, Version 16.2.0, 2020. [Online]. Available: \url{https://portal.3gpp.org/desktopmodules/Specifications/SpecificationDetails.aspx?specificationId=3213}
\BIBentrySTDinterwordspacing

\end{thebibliography}
